\documentclass[final,letterpaper,twocolumn,journal]{IEEEtran}

\usepackage{amsmath,amssymb,graphicx,epstopdf,epsfig}
\usepackage{url,hyperref}
\usepackage[justification=centering]{caption}
\usepackage{lettrine}
\usepackage{color}

\newcommand{\abs}[1]{\left|#1\right|}
\newcommand{\norm}[2]{\left\|#1\right\|_#2}
\newcommand{\normtwo}[1]{\left\|#1\right\|_2}
\newcommand{\normone}[1]{\left\|#1\right\|_1}
\newcommand{\nx}{n_x}
\newcommand{\setX}{{\boldsymbol{\mathcal{X}}}}
\newcommand{\setS}{{\boldsymbol{\mathcal{S}}}}
\newcommand{\vecx}{\boldsymbol{x}}
\newcommand{\vecy}{\boldsymbol{y}}
\newcommand{\vecz}{\boldsymbol{z}}
\newcommand{\vecn}{\boldsymbol{n}}

\newcommand{\matA}{\boldsymbol{A}}
\newcommand{\matI}{\boldsymbol{I}}
\newcommand{\vecxt}{\boldsymbol{\tilde{x}}}
\newcommand{\vecxh}{\boldsymbol{\hat{x}}}
\newcommand{\vech}{\boldsymbol{h}}
\newcommand{\vecho}{\boldsymbol{h_0}}
\newcommand{\veceo}{e_0}

\allowdisplaybreaks

\begin{document}

\title{Optimizing Matrices For Compressed Sensing Using Existing Goodness Measures: Negative Results, And An Alternative}
\author{
	\IEEEauthorblockN{Alankar Kotwal\IEEEauthorrefmark{1} and Ajit Rajwade\IEEEauthorrefmark{2}} \\
	\IEEEauthorblockA{\IEEEauthorrefmark{1}Department of Electrical Engineering,}
	\IEEEauthorblockA{\IEEEauthorrefmark{2}Department of Computer Science and Engineering} \\
	Indian Institute of Technology Bombay \\
	\IEEEauthorblockA{\IEEEauthorrefmark{1}alankar.kotwal@iitb.ac.in,}
	\IEEEauthorblockA{\IEEEauthorrefmark{2}ajitvr@cse.iitb.ac.in} \\
}%

\maketitle

\begin{abstract} \label{sec:abstract}
The bound that arises out of sparse recovery analysis in compressed sensing involves input signal sparsity and some property of the sensing matrix. A directed effort has therefore been made in the literature to optimize the sensing matrices for optimal recovery using this property. We discover, in the specific case of optimizing sensing codes for the CACTI camera \cite{Llull13a}, that the very popular method of mutual coherence minimization does not produce optimal results: codes designed to optimize effective dictionary coherence often perform worse than random codes in terms of mean squared reconstruction error.

This surprising phenomenon leads us to investigate the reliability of the coherence bound for sensing matrix optimization, in terms of its looseness. We examine, on simulated data, the looseness of the bound as it propagates across various steps of the inequalities in a derivation leading to the final bound. We then similarly examine an alternate bound derived in \cite{Tang2015} based on the $\ell_1/\ell_{\infty}$ notion of sparsity, which turns out to be a compromise between coherence and the restricted isometry constant (RIC). Moreover, we also perform a bound looseness analysis for the RIC as in \cite{Cai2010}. The conclusion of these efforts is that coherence optimization is problematic not only because of the coherence bound on the RIC, but also the RIC bound itself. These negative results imply that despite the success of previous work in designing sensing matrices based on optimization of a matrix quality factor, one needs to exercise caution in using them for practical sensing matrix design.

We then introduce an alternative paradigm for optimizing sensing matrices that overcomes the looseness of compressed sensing upper bounds using an average case error approach. We show a proof-of-concept design using this paradigm that performs convincingly better than coherence-based design in the CACTI case, and no worse for general matrices.
\end{abstract}

\begin{IEEEkeywords} \label{sec:keywords}
compressed sensing, sparse recovery, matrix optimization, coherence, bound looseness, worst case errors, average case errors
\end{IEEEkeywords}

\section{Introduction and Related Work} \label{sec:intro}
\lettrine[lines=2]{\scalebox{1}{A}}{} large part of the theoretical development in compressed sensing has dealt with random matrices. This is because matrices drawn from distributions like the Gaussian or the Bernoulli are known to obey some properties such as restricted isometry (RIP) \cite{Foucart2013} with overwhelming probability, making them well suited to compressive recovery. The question to be asked next, therefore, is whether we can do better than random. Do there exist principled and mathematically founded ways to find matrices that are `optimal', in some sense, for recovery using compressed sensing methods?

Here is where the strong theoretical foundations beneath sparse recovery come to our aid. Given a basis that sparsifies the entries in the input signal, we can construct an effective dictionary as a product of the sensing matrix and the basis. The guarantees then express, in terms of some property of the effective dictionary and sparsity of the input signal in the chosen basis, the ability of recovery algorithms to reconstruct the signal from compressive measurements \cite{Studer201412,Candes2008}. The problem is then reduced to finding sensing matrices which recover best, given some error criterion. The traditional error criterion is the $l_2$ error between recovered and true vectors, and the traditional sparsity criterion is the $l_p$ norm ($0 \leq p \leq 1$) of the underlying vector.

The bound that thus arises involves a quantity called the restricted isometry constant (RIC) of a matrix, essentially a measure of how well columns of a matrix can be expressed as sparse combinations of each other. This is a notoriously difficult quantity to calculate: the $s^\text{th}$ RIC requires a listing of $s$-tuples of columns chosen from $n$ columns. In fact computing the RIC is known to be strongly NP-hard \cite{Tillmann2014} and even hard to approximate \cite{Natarajan2014}. This quantity, therefore, is a difficult one to compute. One, therefore, trades complexity off for the tightness of the bound by upper-bounding the RIC using the so-called mutual coherence $\mu$ of a matrix with the Ger\v{s}gorin disc theorem \cite{Studer201412}. The coherence of a $m \times n$ matrix $\boldsymbol{M}$ with columns $\boldsymbol{m_i}$, $\mu(\boldsymbol{M})$, is given as
\begin{equation}
\mu(\boldsymbol{M}) = \max_{i, j \leq n, i \neq j} \frac{\abs{\left< \boldsymbol{m_i}, \boldsymbol{m_j}\right>}}{\sqrt{\abs{\left< \boldsymbol{m_i}, \boldsymbol{m_i}\right> \left< \boldsymbol{m_j}, \boldsymbol{m_j}\right>}}}.
\label{eq:cohDefn}
\end{equation}
This quantity is an efficient one to calculate, and one can even optimize sensing matrices with respect to their mutual coherence value \cite{Duarte200907,Elad200610}. To quote the bound in \cite{Studer201412}, if the compressed sensing problem
\begin{equation}
\vecy = \matA \vecx + \vecn
\label{eq:compSensProblem}
\end{equation}
is solved with the basis pursuit solver
\begin{equation}
\vecxh = \arg \min_{\vecxt} \|\vecxt\|_1 \text{ such that } \|\vecy - \matA\vecxt\|_2 < \epsilon
\label{eq:basisPursuit}
\end{equation}
the following holds: given a particular $s < n$, define $\setX$ as the set of indices of the $s$ absolute greatest entries of $\vecx$. Define the best $s$-sparse approximation to $\vecx$, $\vecx_\setX$, by setting the $\vecx$ values at indices not in $\setX$ to zero. If $\|\vecn\| \leq \eta$ and $$n_x \leq \frac{1}{2} \left(1+\frac{1}{\mu(\matA)}\right)$$ the upper bound on sparse recovery error (as derived in \cite{Studer201412}) in terms of coherence is given as follows:
\begin{equation}
\|\vech\|_2\ \leq C_1 (\epsilon + \eta) + C_2 \|\vecx - \vecx_\setX\|_1
\label{eq:studerBoundMain}
\end{equation}
It can be verified that both the coefficients $C_1$ and $C_2$ are increasing functions of the mutual coherence.

One of the earliest efforts to address the matrix design question \cite{Elad200610}, therefore, involves a strategy that minimized coherence as a function of effective dictionary $\mathbf{D}$ entries. The approach taken involves constructing a Gram matrix $\mathbf{G} = \mathbf{D}^T \mathbf{D}$, applying a non-linear transformation to the off-diagonal elements to decrease the magnitude of dot products between columns of $\mathbf{D}$, and attempting to extract a new effective dictionary from the new Gram matrix. Another method \cite{Duarte200907} involved a design of sensing matrix entries by minimizing the Frobenius norm of the departure of the Gram matrix of the effective dictionary from the identity matrix. There have been a plethora of other efforts for compressed sensing design since, all using the coherence as a goodness criterion for sensing matrices. For instance, \cite{Mordechay2014} designs an optimal energy-preserving sensing matrix for Poisson compressed sensing, where the optimizing criterion is the coherence directly. \cite{Abolghasemi10} uses a method similar to \cite{Duarte200907} for optimizing general sensing matrices for coherence with gradient descent. In \cite{Pereira14} is a method to design sensing matrices maximally incoherent with the sparsifying orthogonal basis. \cite{Parada17} applies coherence minimization to design structured matrices for the Coded Aperture Snapshot Spectral Imaging (CASSI) system \cite{Gehm07,Wagadarikar08}. \cite{Bouchhima15} and \cite{Obermeier17} apply coherence-based design to environmental sounds and electromagnetic compressed sensing applications respectively. 

Motivated by this, we attempt to incorporate the structure imposed by the acquisition framework in the Coded Aperture Compressive Temporal Imaging (CACTI) system \cite{Llull13a, Llull13b} into coherence-based design. The sensing matrix in the CACTI system is essentially derived from a 2D mask pattern which can be shifted in time. Therefore, one seeks to find a mask pattern that yields least mutual coherence, given a specific sparsifying basis. To our surprise, we observe that such a procedure fails to produce any reasonable improvement in the reconstruction quality afforded by the matrix created from a designed mask pattern over a matrix created from a random mask pattern. This sets us into an exploration of why this happens, and we discover that the derived compressed sensing bounds, though known to be loose, are loose enough to allow a huge margin of error. Evaluating quantities intermediate in the derivation of these bounds corroborates our beliefs and provides examples of situations when optimizing an upper bound on a quantity does not necessarily optimize the quantity itself. A realization that the coherence is a loose worst case bound then leads us to investigate if the average case error, as well as other criteria such as RIC, are better as optimization criteria.

\emph{Organization of the paper}: The rest of this paper is organized as follows:
\begin{itemize}
    \item Section~\ref{sec:cacti} introduces the sensing model for the CACTI camera from \cite{Llull13a, Llull13b}, describes our image model and scheme for optimizing codes for it. It also shows negative results from the optimization, and shows a counterexample that minimizing upper bounds leads to minimizing lower bounds;
    \item Section~\ref{sec:bound} then analyzes the sparse recovery bound in \cite{Studer201412} (quoted above in Eq.~\ref{eq:studerBoundMain}) empirically, in terms of looseness propagation across the inequalities in the derivation and provides a discussion of the results;
    \item Section~\ref{sec:newBound} explores an alternate, tractable error bound derived in \cite{Tang2015} for compressed sensing recovery and performs a similar looseness analysis between the bound and actual error over a dataset of vectors;
    \item Section~\ref{sec:ric} analyzes looseness of bounds based on a newly introduced matrix quality measure based on the $\ell_1/\ell_{\infty}$ notion of sparsity  \cite{Tang2015}, which is a compromise between RIC and coherence. It then takes a look at how loose the RIC bound is for reasonably sized matrices at small sparsity levels, where the bound in \cite{Cai2010} applies;
    \item Section~\ref{sec:mmse} motivates the average squared error over the distribution of input vectors as a possible optimization criterion and shows a proof-of-concept design and reconstruction results from this design. These results demonstrate the advantages of the mean squared error as an optimization criterion over coherence;
    \item Section~\ref{sec:conclusion} summarizes the findings in this paper, and provides a potential direction for future work in optimizing sensing matrices by deriving average case error bounds.
\end{itemize}

\section{The CACTI Camera} \label{sec:cacti}
The principal idea behind the design of the CACTI camera, in the words of the authors of \cite{Llull13a} is using ``mechanical translation of a coded aperture for code division multiple access (CDMA) compression of video''. The setup is shown in Fig. 1 of \cite{Llull13a}. The camera achieves compression across time by combining frames into coded snapshots while sensing and separating them during reconstruction. $T$ input frames of size $N_1 \times N_2$ denoted as $\{\boldsymbol{X_i}\}_{i=1}^{T}$ are sensed so that output $\boldsymbol{Y}$ of size $N_1 \times N_2$ appears as a pixel-wise coded superposition (dictated by the sensing matrices $\boldsymbol{\Phi_i}$) of the input frames. The sensing framework is
\begin{equation}
\textrm{vec}(\boldsymbol{Y}) = \sum_{i=1}^{T} \boldsymbol{\Phi_i} \textrm{vec}(\boldsymbol{X_i}) = \boldsymbol{\Phi} \textrm{vec} (\boldsymbol{X}).
\label{eq:cactiSensing}
\end{equation}
Here, each $\boldsymbol{\Phi_i}$ is a $N_1 N_2 \times N_1 N_2$ non-negative (possibly, but not necessarily binary) diagonal sensing matrix with the vectorized code elements on the diagonal, and the overall sensing matrix $\boldsymbol{\Phi} = (\boldsymbol{\Phi_1} | \boldsymbol{\Phi_2} | ... \boldsymbol{\Phi_T})$ has size $N_1 N_2 \times N_1 N_2 T$. The complete $N_1 \times N_2 \times T$ video is represented as $\boldsymbol{X} = (\boldsymbol{X_1} | \boldsymbol{X_2} | ... | \boldsymbol{X_T})$. The sensing matrices $\boldsymbol{\Phi_i}$ are not independent across $i$: the mechanical translation amounts to a fixed circular shift in the diagonal elements of $\boldsymbol{\Phi_i}$, dictated by the set mechanical translations. Because of this relationship, there is a fixed set of $N_1N_2$ values used as code elements. Let us define a vector $\boldsymbol{\Phi}$ of dimension $N_1N_2 \times 1$ to be the vector of the values in this set. Let the $j^\text{th}$ diagonal element of $\boldsymbol{\Phi_i}$ be $\Phi_{ij}$ and that of $\boldsymbol{\Phi}$ be $\Phi^j$. 

It is now of importance to note where each $\Phi_{ij}$ came from. The $i^\text{th}$ mechanical permutation takes $\boldsymbol{\Phi}$ to $\boldsymbol{\Phi_i}$. Clearly, if we apply the permutation $i$ to the vector $(1\ 2\ 3\ \hdots\ n)^T$ and call the resultant vector $\boldsymbol{p_i}$, the $j^\text{th}$ element of $\boldsymbol{p_i}$ will denote which element of $\boldsymbol{\Phi}$ got circularly shifted to $\Phi_{ij}$. Defining $p_{ij}$ to be the $j^\text{th}$ element of $\boldsymbol{p_i}$, we have $\Phi_{ij} = \Phi^{p_{ij}}$. 

We use a 2D-DCT basis $\boldsymbol{D}$ to model each frame in the input data, though our method is general enough to work for any other basis. The dictionary $\boldsymbol{\Psi}$ sparsifying the entire video sequence, thus, is a block-diagonal matrix with the $n \times n$ sparsifying basis $\boldsymbol{D}$ on the diagonal where $n = N_1 N_2$ is the number of pixels per video frame. Thus,
\begin{align}
\textrm{vec}(\boldsymbol{Y}) &= \begin{pmatrix}
\boldsymbol{\Phi_1} & \hdots & \boldsymbol{\Phi_T}
\end{pmatrix}
\begin{pmatrix}
\boldsymbol{D \alpha_1} &
\hdots &
\boldsymbol{D \alpha_T}
\end{pmatrix}^T \\
&= \begin{pmatrix}
\boldsymbol{\Phi_1 D} & \hdots & \boldsymbol{\Phi_T D}
\end{pmatrix}
\begin{pmatrix}
\boldsymbol{\alpha_1} &
\hdots &
\boldsymbol{\alpha_T}
\end{pmatrix}^T
\label{eq:cactiModel}
\end{align}

\noindent Given a measurement $\boldsymbol{Y}$, we recover the input $\{\boldsymbol{X_i}\}_{i=1}^{T}$ through the DCT coefficients $\boldsymbol{\alpha}$ by solving the optimization problem
\begin{equation}
\begin{split}
\min_{\boldsymbol{\alpha}} \|\boldsymbol{\alpha}\|_1 \text{ subject to } \\ \textrm{vec}(\boldsymbol{Y}) = \boldsymbol{\tilde{\Phi} \Psi \alpha},\\ \boldsymbol{\alpha} = \begin{pmatrix}
\boldsymbol{\alpha_1} &
\boldsymbol{\alpha_2} &
\hdots &
\boldsymbol{\alpha_T}
\end{pmatrix}^T \\
\boldsymbol{\tilde{\Phi}} = \begin{pmatrix}
\boldsymbol{\Phi_1} & \hdots & \boldsymbol{\Phi_T}
\end{pmatrix}
\end{split}
\label{eq:cactiRecOpt}
\end{equation}
In our implementation we used the \texttt{CVX}~\cite{cvx} solver for solving the convex optimization problem in Eq.~\ref{eq:cactiRecOpt}.

\subsection{Optimizing for the CACTI camera} \label{subsec:cactiOpt}
We follow an explicit coherence minimization policy similar to \cite{Full} for optimizing codes. To minimize coherence, we write down the expression for the coherence of the joint dictionary $\boldsymbol{\tilde{\Phi} \Psi}$:
\begin{align}
\boldsymbol{\tilde{\Phi} \Psi} = \begin{pmatrix}
\boldsymbol{\Phi_1} \boldsymbol{D} & \boldsymbol{\Phi_2} \boldsymbol{D} & \hdots & \boldsymbol{\Phi_T} \boldsymbol{D}
\end{pmatrix}.
\end{align}

Define the $\mu^\text{th}$ column of $\boldsymbol{D^T}$ to be $\boldsymbol{d_\mu}$, and its $\beta^\text{th}$ element as $d_{\mu}\left(\beta\right)$. Let the variables $\mu$ and $\nu$ go from 1 to $n$, and the variables $\beta$ and $\gamma$ go from 1 to $T$. In a similar way to \cite{Full} (see appendix in \cite{Full}), using the steps outlined in the appendices, we write the normalized dot product between the $\beta^\text{th}$ column of the $\mu^\text{th}$ block of the effective dictionary and the $\gamma^\text{th}$ column of the $\nu^\text{th}$ block of the effective dictionary as
\begin{align}
M_{\mu \nu}(\beta\gamma) &= \frac{\sum_{\alpha = 1}^{n} \Phi_{\mu \alpha} \Phi_{\nu \alpha} d_\alpha (\beta) d_\alpha (\gamma)}{\sqrt{\left( \sum_{\alpha = 1}^{n} \Phi_{\mu \alpha}^2 d^2_\alpha (\beta) \right) \left( \sum_{\tau = 1}^{n} \Phi_{\nu \tau}^2 d^2_\tau (\gamma) \right)}} \\
&= \frac{\sum_{\alpha = 1}^{n} \Phi^{p_{\mu \alpha}} \Phi^{p_{\nu \alpha}} d_\alpha (\beta) d_\alpha (\gamma)}{\sqrt{\left( \sum_{\alpha = 1}^{n} \Phi^{p_{\mu \alpha}2} d^2_\alpha (\beta) \right) \left( \sum_{\tau = 1}^{n} \Phi^{p_{\nu \tau}2} d^2_\tau (\gamma) \right)}}.
\end{align}

With the numerator of the above expression renamed to $\chi_{\mu \nu}(\beta\gamma)$ and the denominator renamed to $\xi_{\mu \nu}(\beta\gamma)$, we write, 
\begin{equation}
\begin{split}
\frac{d\chi_{\mu \nu}(\beta\gamma)}{d\Phi_{\delta \epsilon}} &= d_\epsilon (\beta) d_\epsilon (\gamma) \left( \Phi_{\mu \epsilon} \uparrow_{\nu \delta} + \uparrow_{\mu \delta} \Phi_{\nu \epsilon} \right) \\
\implies \frac{d\chi_{\mu \nu}(\beta\gamma)}{d\Phi^{p_{\delta \epsilon}}} &= d_\epsilon (\beta) d_\epsilon (\gamma) \left( \Phi^{p_{\mu \epsilon}} \uparrow_{\nu \delta} + \uparrow_{\mu \delta} \Phi^{p_{\nu \epsilon}} \right).
\end{split}
\end{equation}

\begin{equation}
\begin{split}
\frac{d\xi_{\mu \nu}(\beta\gamma)}{d\Phi_{\delta \epsilon}} &= \frac{1}{\xi_{\mu \nu}(\beta \gamma)} \left[ \Phi_{\mu \epsilon} d_\epsilon^2(\beta) \uparrow_{\mu \delta} \sum_{\tau = 1}^{n} \Phi_{\nu \tau}^2 d^2_\tau (\gamma) \right. + \\
&\qquad \qquad \left. \Phi_{\nu \epsilon} d_\epsilon^2(\gamma) \uparrow_{\nu \delta} \sum_{\alpha = 1}^{n} \Phi_{\mu \alpha}^2 d^2_\alpha (\beta) \right] \\
\implies \frac{d\xi_{\mu \nu}(\beta\gamma)}{d\Phi^{p_{\delta \epsilon}}} &= \frac{1}{\xi_{\mu \nu}(\beta \gamma)} \left[ \Phi^{p_{\mu \epsilon}} d_\epsilon^2(\beta) \uparrow_{\mu \delta} \sum_{\tau = 1}^{n} \Phi^{p_{\nu \tau}2} d^2_\tau (\gamma) + \right. \\ 
&\qquad \qquad \left. \Phi^{p_{\nu \epsilon}} d_\epsilon^2(\gamma) \uparrow_{\nu \delta} \sum_{\alpha = 1}^{n} \Phi^{p_{\mu \alpha}2} d^2_\alpha (\beta) \right].
\end{split}
\end{equation}

We perform a projected (to maintain non-negativity of $\boldsymbol{\Phi}$) gradient descent with adaptive step-size and use a multi-start strategy to combat the non-convexity of the problem.

\subsection{Results on simulated data} \label{subsec:cactiRes}
\begin{figure}[!h]
\centering
\includegraphics[scale=0.2]{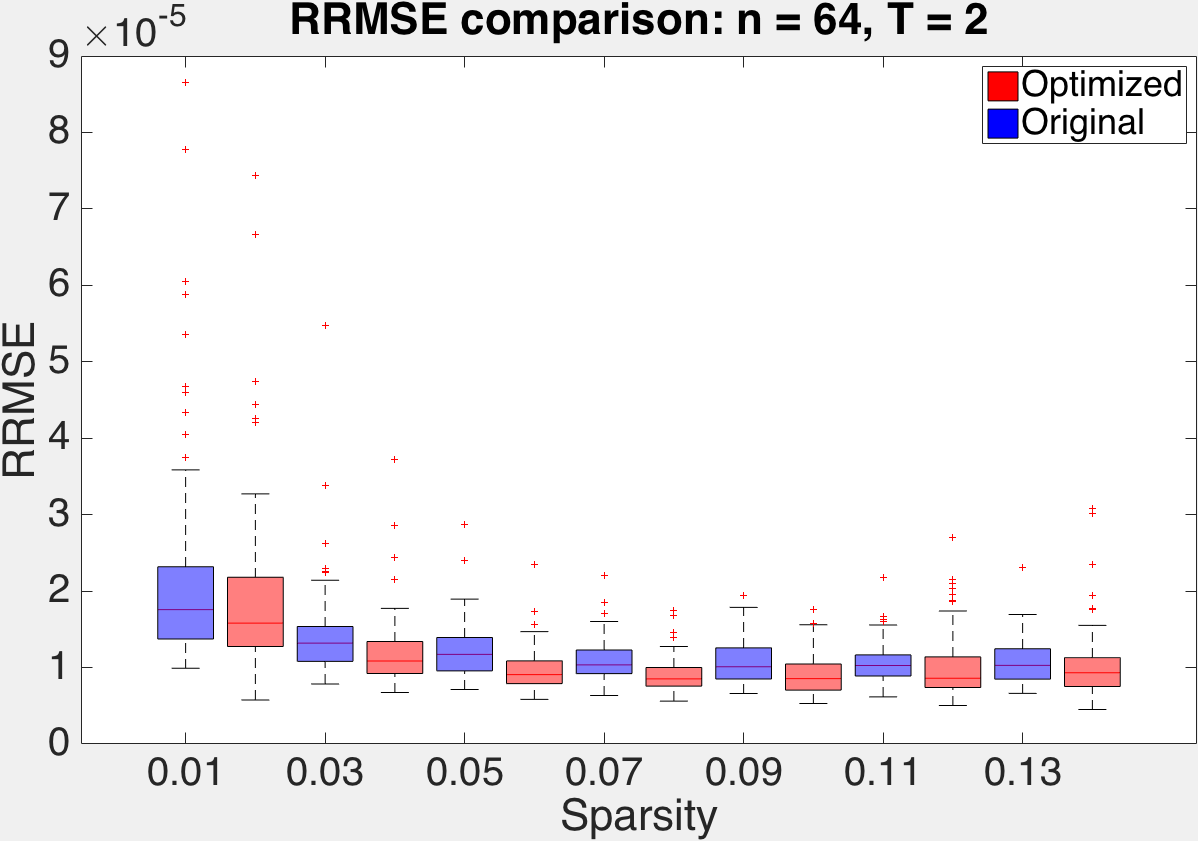}
\caption{Average RRMSE as a function of sparsity for $8 \times 8$ signals, sparse in 2D DCT, combined with $T = 2$. Permutations: [5, 3; 6, 8].}
\label{fig:cactiRRMSE2}
\end{figure}

\begin{figure}[!h]
\centering
\includegraphics[scale=0.2]{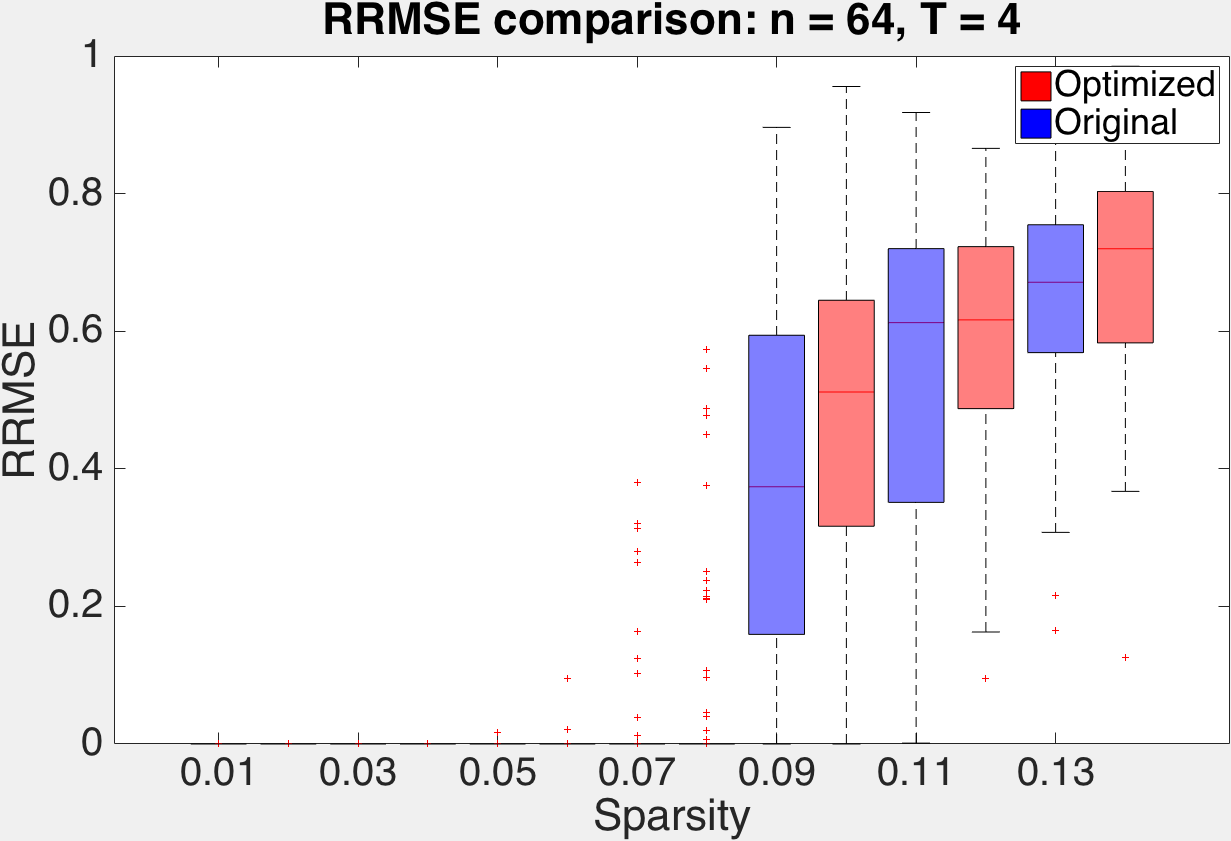}
\caption{Average RRMSE as a function of sparsity for $8 \times 8$ signals, sparse in 2D DCT, combined with $T = 4$. Permutations: [7, 8; 2, 8; 6, 1; 3, 5]}
\label{fig:cactiRRMSE4}
\end{figure}

\begin{figure}[!h]
\centering
\includegraphics[scale=0.2]{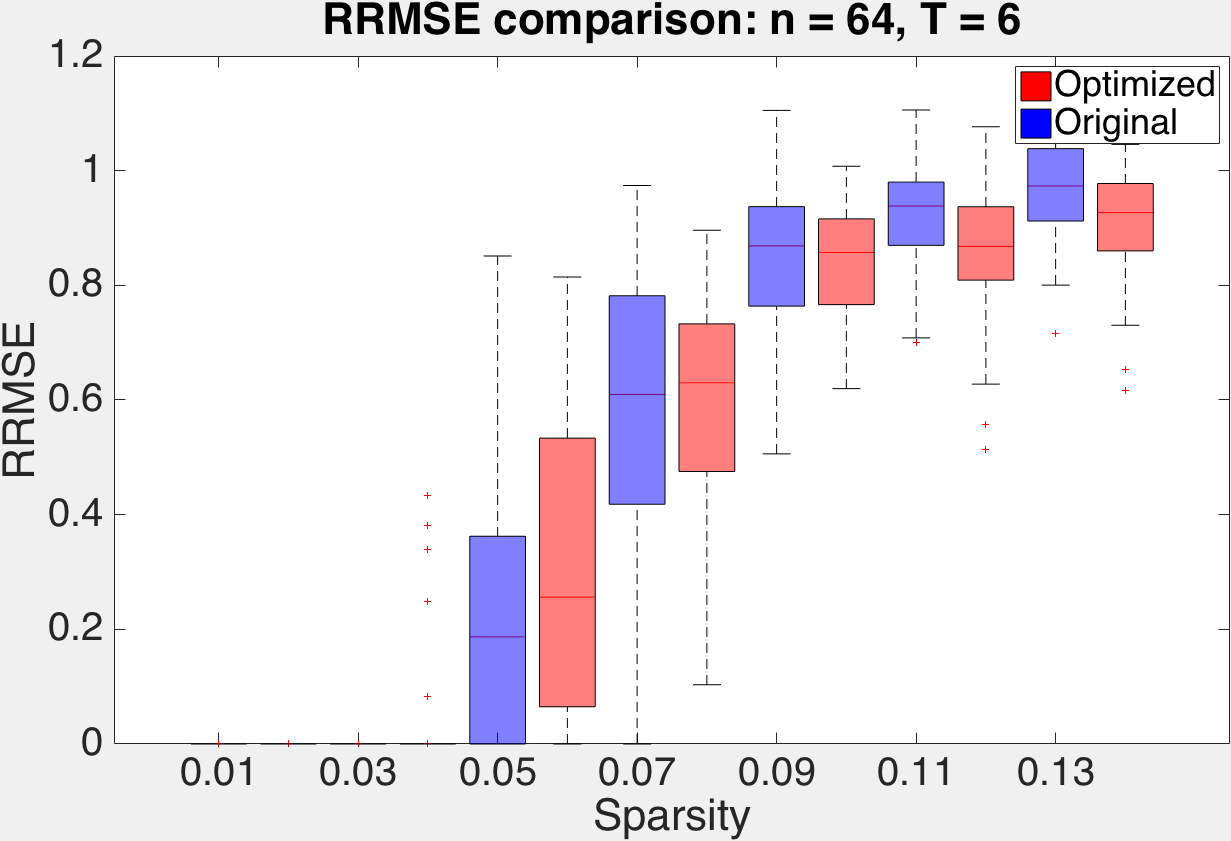}
\caption{Average RRMSE as a function of sparsity for $8 \times 8$ signals, sparse in 2D DCT, combined with $T = 6$. Permutations: [6, 7; 3, 6; 6, 2; 1, 4; 8, 3; 5, 2]}
\label{fig:cactiRRMSE6}
\end{figure}

Here, we experiment with toy data where we can precisely control the sparsity of the input signals. Specifically, assuming a set of mechanical translations, we randomly generate $T$ $s$-sparse (in 2D DCT) $8 \times 8$ signals $\{\boldsymbol{X_i}\}_{i=1}^{T}$, combine them as per the sensing framework in Eq.~\ref{eq:cactiSensing}, using matrices formed by using random codes and our designed codes as per the structure in Eq.~\ref{eq:cactiModel} and add noise bounded in norm to $\epsilon=10^{-5}$ to get $\boldsymbol{Y}$. Average RRMSE errors on doing this over a set of 100 vectors, as a function of signal sparsity and compression level $T$ are shown in Figs. \ref{fig:cactiRRMSE2}, \ref{fig:cactiRRMSE4}, and \ref{fig:cactiRRMSE6}.

These figures don't tell a very happy story: the optimization technique completely fails to produce any statistically significant improvement in the error over random matrices. The coherence decreases are significant: in the best case, the coherence for $T = 2$ decreases from 0.7911 to 0.3462, for $T = 4$ decreases from 0.7921 to 0.4952, and for $T = 6$ decreases from 0.9156 to 0.5430. 

What causes the algorithm to fail? The underlying assumption in this method is that the bound that RIC establishes on the maximum recovery error surface plotted against the space of sensing matrices behaves close to the actual maximum error surface (coherence further loosens up the bound). However, this might not be the case: the looser the bound gets, the more freedom the error surface has to deviate from the behavior of the bound. Then, minimizing the maximum in the bound may not correspond to minimizing the maximum in the actual error surface.

\begin{figure}[!h]
\centering
\includegraphics[scale=0.2]{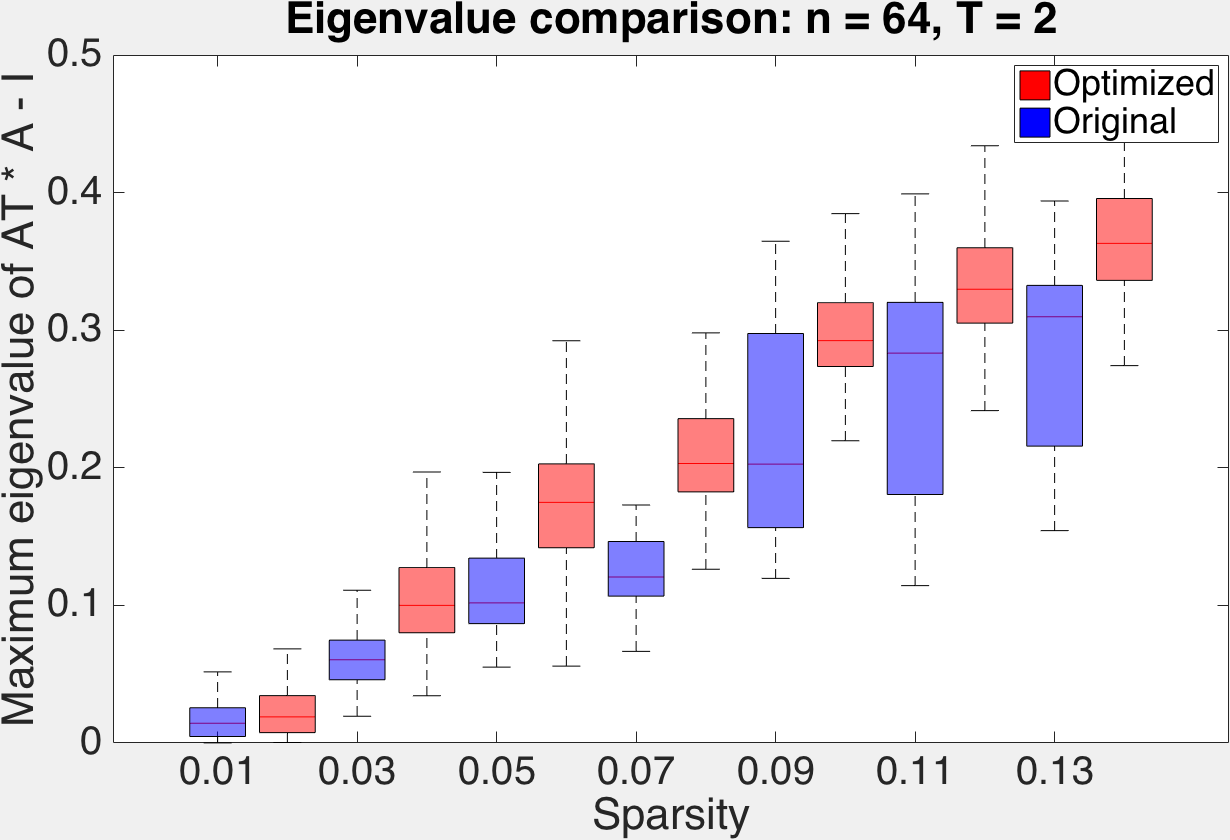}
\caption{Absolute maximum restricted eigenvalue boxplot as a function of sparsity for $8 \times 8$ signals, sparse in 2D DCT, with $T = 2$. Permutations: [5, 3; 6, 8]}
\label{fig:cactiEigen2}
\end{figure}

\begin{figure}[!h]
\centering
\includegraphics[scale=0.2]{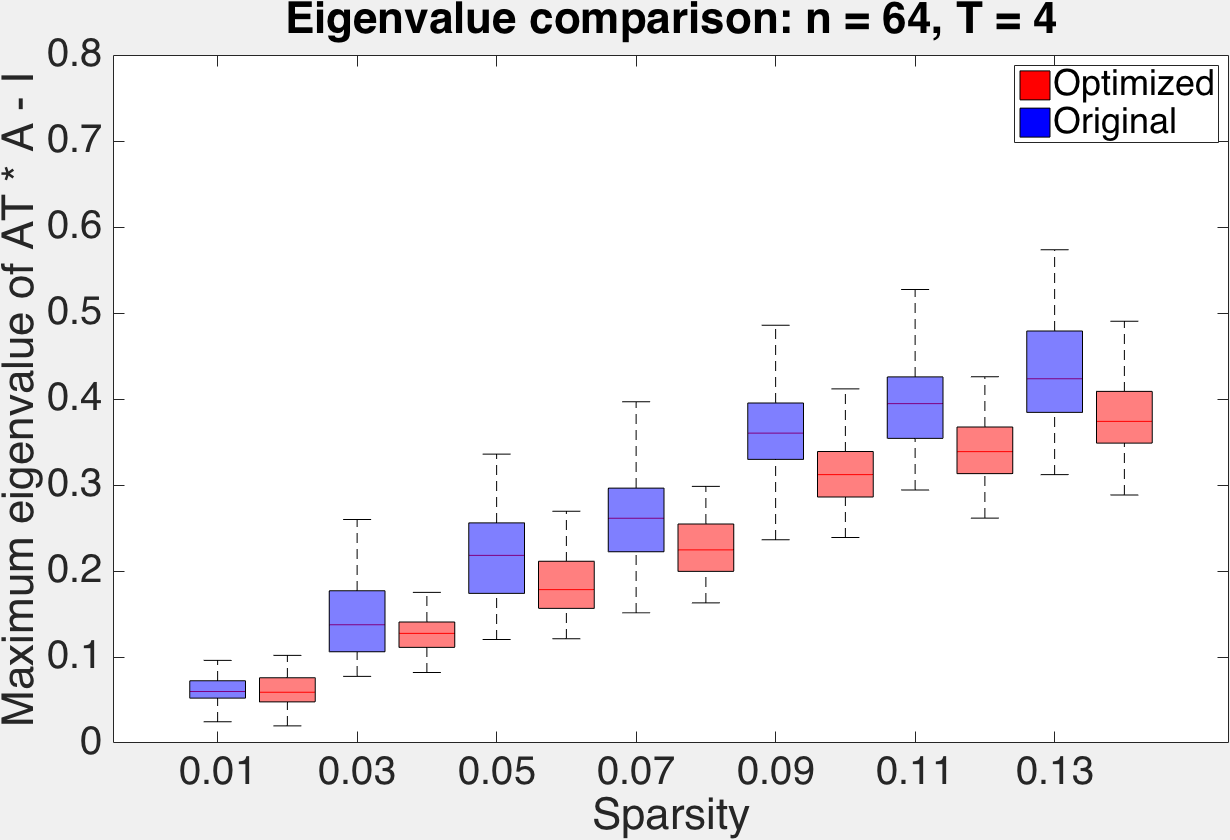}
\caption{Absolute maximum restricted eigenvalue boxplot as a function of sparsity for $8 \times 8$ signals, sparse in 2D DCT, with $T = 4$. Permutations: [7, 8; 2, 8; 6, 1; 3, 5]}
\label{fig:cactiEigen4}
\end{figure}

\begin{figure}[!h]
\centering
\includegraphics[scale=0.2]{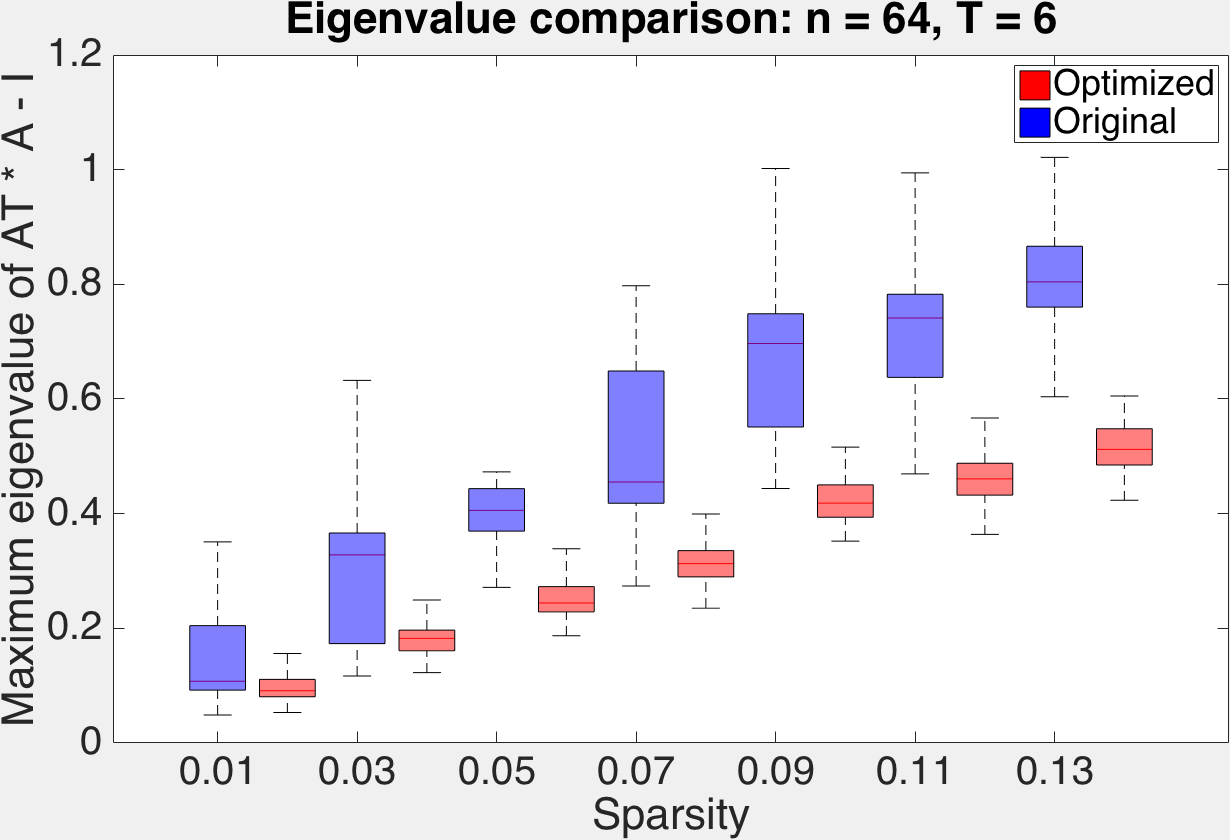}
\caption{Absolute maximum restricted eigenvalue boxplot as a function of sparsity for $8 \times 8$ signals, sparse in 2D DCT, with $T = 6$. Permutations: [6, 7; 3, 6; 6, 2; 1, 4; 8, 3; 5, 2]}
\label{fig:cactiEigen6}
\end{figure}

To quantify this concept, suppose that some oracle gives us the actual supports of the vectors in the dataset. Then, a pseudoinverse solution is possible at sufficiently small sparsities. Defining $\lambda_k\left(\boldsymbol{M}\right)$ as the $k^\text{th}$ largest absolute eigenvalue of the matrix $M$, note that the $s^\text{th}$ RIC of a matrix $\boldsymbol{A} \triangleq \boldsymbol{\tilde{\Phi} \Psi}$ is the following:
\begin{equation}
\delta_s = \max_{\setS \in \{1\ ...\ n\}} |\lambda_1(\matA_\setS^T \matA_\setS - \matI)|.
\end{equation}
This is the maximum vector-induced 2-norm of the matrix $\matA_\setS^T \matA_\setS - \matI$ over all subsets $\setS$ of $\{1,2,...,n\}$ with size $s$, where $\boldsymbol{A_{\setS}}$ is a submatrix of $\boldsymbol{A}$ with columns from set $\setS$. Each $\matA_\setS^T \matA_\setS - \matI$ appears in the error bound for recovery for a vector with known support $\setS$. We therefore plot, for the entire dataset of vectors we used to make the RRMSE plot, the absolute maximum eigenvalue of $\matA_\setS^T \matA_\setS - \matI$, where $\setS$ is the support of the vector, for both random $\matA$ of the form imposed by CACTI, and designed $\matA$. The error in reconstructing this particular vector is bounded tighter than coherence by this absolute maximum eigenvalue. This gives us a handle on how well minimizing coherence minimizes this eigenvalue across supports, and therefore how much we lose by relaxing the RIC to coherence. The maximum eigenvalues for random CACTI and design CACTI matrices for $T = 2,4,6$ are shown in Figs. \ref{fig:cactiEigen2}, \ref{fig:cactiEigen4} and \ref{fig:cactiEigen6} respectively, for randomly chosen permutations. No significant difference was observed for other sets of permutations.

The $T = 2$ case is surprising: decreasing coherence starting from random matrices seems to increase the values of the absolute maximum eigenvalues, which goes directly against the assumption involved in minimizing coherence. The $T = 4$ and $T = 6$ cases behave better in terms of eigenvalues, though their performance in terms of RRMSE error isn't very good. These findings point to the fact that the problem lies not only in the relaxation of the RIC to the coherence, but also in the RIC bound itself. 

\textbf{Using Average Coherence:} A tempting thought, at this juncture, is to maximize some average of the squares of the dot products of pairs of (non-identical) columns in $\boldsymbol{A}$, instead of the coherence, which is the maximum of these. This is because the coherence bound on the RIC relaxes the sum of $k-1$ off-diagonal elements in $\boldsymbol{A}^T \boldsymbol{A}$ to $k-1$ times the maximum, which is the coherence. Designing matrices optimizing the square of these off-diagonal elements and performing simulated data experiments similar to the above produce similar RRMSE behavior: the matrices designed this way are no better than matrices designed using just coherence, and certainly no better than random matrices. Besides, this is a heuristic approach because no theoretical reconstruction guarantees have been derived for such a measure of average coherence in the compressed sensing literature yet. 

This warrants a more detailed empirical understanding of error bounds in compressed sensing, focusing on how bounds evolve across inequalities that give rise to them. This will be the subject of the next section.

\section{The compressed sensing bound} \label{sec:bound}
The bound we choose to examine is the one quoted in Eq.~\ref{eq:studerBoundMain} for recovery of nearly sparse vectors using basis pursuit. Let the compressed sensing scenario be as in Eq.~\ref{eq:compSensProblem}, with an overcomplete $m \times n$-sized $\matA$, $n \times 1$-sized $\vecx$ and $m \times 1$-sized $\vecy$, recovering $\vecxh$ by solving the basis pursuit problem in Eq.~\ref{eq:basisPursuit}.

We will quote some relevant steps from the proof of the bound in Eq.~\ref{eq:studerBoundMain}, following \cite{Studer201412}. Let $\vech = \vecxh - \vecx$. Construct $\vecho = \vech_\setX$ by setting elements of $\vech$ at indices in $\setX^C$ to zero. Let $\veceo = 2 \|\vecx - \vecx_\setX\|_1$. Then, using the above definitions,
\begin{align}
\|\vecx\|_1 &\geq \|\vecxh\|_1 \\
&= \|\vecxh_\setX\|_1 + \|\vecxh_{\setX^C}\|_1 \\ 
&= \|\vecx_\setX + \vecho\|_1 + \|\vech-\vecho + \vecx_{\setX^C}\|_1 \\
&\geq \|\vecx_\setX\|_1 - \|\vecho\|_1 + \|\vech-\vecho\|_1 - \|\vecx_{\setX^C}\|_1 \\
&\implies \|\vech-\vecho\|_1 \leq 2 \|\vecx_{\setX^C}\|_1 + \|\vecho\|_1 \\ 
&\implies \|\vech-\vecho\|_1 \leq \|\vecho\|_1 + \veceo \label{eq:coneconstraint}\\
&\implies \|\vech\|_1 \leq 2 \|\vecho\|_1 + \veceo 
\end{align}
where the last step follows from the reverse triangle inequality. Furthermore, 
\begin{align}
\|\matA \vech\|_2 &= \|\matA \vecxh - \vecy - (\matA \vecx - \vecy)\|_2 \\
&\leq \|\matA \vecxh - \vecy\|_2 + \|\matA \vecx - \vecy\|_2 \\
&\leq \eta + \epsilon \label{eq:tubeconstraint}
\end{align}
An application of the Ger\v{s}gorin disc theorem to $\|\matA \vecho\|^2$ gives, since $\vecho$ is perfectly sparse
\begin{align}
(1 - \mu (n_x - 1)) \|\vecho\|_2^2 \leq \|\matA\vecho\|_2^2 \leq (1 + \mu (n_x - 1)) \|\vecho\|_2^2
\label{eq:gersgorindiscthm}
\end{align}
Next,
\begin{align}
\abs{\vech^T \matA^T \matA \vecho} & \geq \abs{\vecho^T \matA^T \matA \vecho} - \abs{(\vech-\vecho)^T \matA^T \matA \vecho} \label{eq:diffs4n4} \\
& \geq \left(1-\mu(\nx-1)\right)\normtwo{\vecho}^2 \nonumber \\ & \qquad \qquad - \abs{\sum_{k\in\setX}\sum_{l\in\setX^c} [\vecho^T]_k \boldsymbol{a_k}^T \boldsymbol{a_l} [\vech]_l} \label{eq:errbound0} \\
& \geq \left(1-\mu(\nx-1)\right)\normtwo{\vecho}^2 \nonumber \\ & \qquad \qquad \qquad - \mu\normone{\vecho}\normone{\vecho} \label{eq:errbound1} \\
& \geq \left(1-\mu(\nx-1)\right)\normtwo{\vecho}^2 \nonumber \\ & \qquad \qquad \qquad - \mu\normone{\vecho}\left(\normone{\vecho} + \veceo\right) \label{eq:errbound2}\\
& \geq \left(1-\mu(\nx-1)\right)\normtwo{\vecho}^2 - \mu \nx \normtwo{\vecho}^2 \nonumber \\ & \qquad \qquad \qquad - \mu\sqrt{\nx}\normtwo{\vecho} \veceo \label{eq:errbound3} \\
& = \left(1-\mu(2\nx-1)\right)\normtwo{\vecho}^2 \nonumber \\ & \qquad \qquad \qquad - \mu\sqrt{\nx}\normtwo{\vecho} \veceo, \label{eq:errbound4}
\end{align}
Eq. \ref{eq:errbound0} is a result of the Ger\v{s}gorin bound in Eq. \ref{eq:gersgorindiscthm}, Eq. \ref{eq:errbound1} arises from $\abs{\boldsymbol{a_k}^T\boldsymbol{a_l}} \leq \mu$,  $\forall\ k\neq l$ and Eq. \ref{eq:errbound2} comes from the condition in Eq. \ref{eq:coneconstraint}. Eq. \ref{eq:errbound3} is an application of the Cauchy-Schwarz inequality. Next,
\begin{align}
\normtwo{\vecho} &\leq \frac{\abs{\vech^T\matA^T\matA\vecho} + \mu\sqrt{\nx}\normtwo{\vecho}\veceo}{\left(1-\mu(2\nx-1)\right)\normtwo{\vecho}} \label{eq:errbound5} \\
& \leq \frac{\normtwo{\matA\vech}\normtwo{\matA\vecho}+\mu\sqrt{\nx}\normtwo{\vecho}\veceo}{\left(1-\mu(2\nx-1)\right)\normtwo {\vecho}} \label{eq:errbound6} \\
& \leq \frac{(\epsilon+\eta) \sqrt{1+\mu(\nx-1)}\normtwo{\vecho} + \mu\sqrt{\nx}\normtwo{\vecho}\veceo}{\left(1-\mu(2\nx-1)\right)\normtwo{\vecho}} \label{eq:errbound7} \\
& = \frac{ (\epsilon+\eta) \sqrt{1+\mu(\nx-1)}+\mu\sqrt{\nx}\veceo}{ 1-\mu(2\nx-1)} \label{eq:errbound7b}
\end{align}

The proof further goes on to bound $\normtwo{\vech}$. The rest of the proof, however, is an application of the bound yet derived, and we will study the looseness of just this part of the proof. For completeness, we quote the bound here.
\begin{align}
\|\vech\|_2\ \leq\ &\frac{1 - \mu (2 n_x - 1) + \sqrt{\mu n_x} \sqrt{1 + \mu (n_x - 1)}}{\sqrt{1 + \mu (2 n_x - 1)}}\ (\epsilon + \eta)\ + \label{eq:finalBound} \\
&\frac{2\sqrt{\mu + \mu^2}}{1 - \mu(2 n_x - 1)}\ \|\vecx - \vecx_\setX\|_1 \notag
\end{align}

\subsection{Important steps in the bound} \label{subsec:steps}
We will concentrate on the simplest case, where the vector $\vecx$ is exactly sparse, and set $n_x = \|\vecx\|_0$. $\|\vecx\|_0$ is set as the maximum $l_0$ norm that the bound allows, which is the greatest integer below $0.5(1 + 1/\mu)$. The definitions then reduce $e_0$ to 0. We generate a set of positive sparse vectors of a size suitable to be sensed with a selected sensing matrix, add noise bounded in norm by $\eta = \epsilon = 10^{-5}$, and plot a boxplot of the relative difference between the left and right hand sides in selected inequalities in the proof above. That means, if an inequality such as $a \leq b$ exists in the bound, and we want to show a relative difference with respect to the left hand side, we show a boxplot, computed over a dataset of our randomly drawn vectors $\vecx_i$, the following quantity:
\begin{equation}
\textrm{Relative Difference}(a,b) = \abs{b-a}/\abs{a}.
\end{equation}

The relative differences we choose to show are due to, in order,
\begin{enumerate}
\item The triangle inequality in Eq. \ref{eq:diffs4n4}, with respect to the left hand side of Eq. \ref{eq:diffs4n4}  
\item The Ger\v{s}gorin bound and triangle inequality in Eq. \ref{eq:errbound0}, with respect to the left hand side of Eq. \ref{eq:diffs4n4}
\item Replacing dot products with their maximum, coherence, in Eq. \ref{eq:errbound1}, with respect to the left hand side of Eq. \ref{eq:diffs4n4} \label{chk:step3}
\item The application of Eq. \ref{eq:coneconstraint} in Eq. \ref{eq:errbound2}, with respect to the left hand side of Eq. \ref{eq:diffs4n4} \label{chk:step4}
\item The bound relating the $l_1$ and $l_2$ norms in Eq. \ref{eq:errbound4}, with respect to the left hand side of Eq. \ref{eq:diffs4n4} \label{chk:step5}
\item The rearrangement of Eq. \ref{eq:errbound5}, with respect to the left hand side of Eq. \ref{eq:errbound5} \label{chk:step6}
\item The Cauchy-Swartz inequality in Eq. \ref{eq:errbound6}, with respect to the left hand side of Eq. \ref{eq:errbound4} \label{chk:step7}
\item The application of Eq. \ref{eq:coneconstraint} in Eq. \ref{eq:errbound7b}, with respect to the left hand side of Eq. \ref{eq:errbound4} \label{chk:step8}
\item The leftmost side of the Ger\v{s}gorin bound in Eq. \ref{eq:gersgorindiscthm}, with respect to $\|\matA \vecho\|_2^2$ \label{chk:stepGnG1} 
\item The rightmost side of the Ger\v{s}gorin bound in Eq. \ref{eq:gersgorindiscthm}, with respect to $\|\matA \vecho\|_2^2$ \label{chk:stepGnG2} 
\item The actual RRMSE error for the simulated vector in question, which is the left hand side of Eq. \ref{eq:finalBound}, and the bound predicted by Eq. \ref{eq:finalBound}, with respect to the actual RRMSE.
\end{enumerate}

\subsection{Looseness propagation} \label{subsec:looseness}
\subsubsection{General sensing matrices} \label{subsubsec:generalLooseness}
Here, to allow a high $\|\vecx\|_0$, we first choose a random $499 \times 500$ matrix drawn from a Gaussian distribution. Also as a benchmark for a compressive scenario, we test with $250 \times 500$ and $125 \times 500$ matrices corresponding to 50\% and 25\% measurement fractions. All this is done at a noise level and reconstruction tolerance of $\epsilon = 10^{-5}$.

\begin{figure}[!h]
\centering
\includegraphics[scale=0.2]{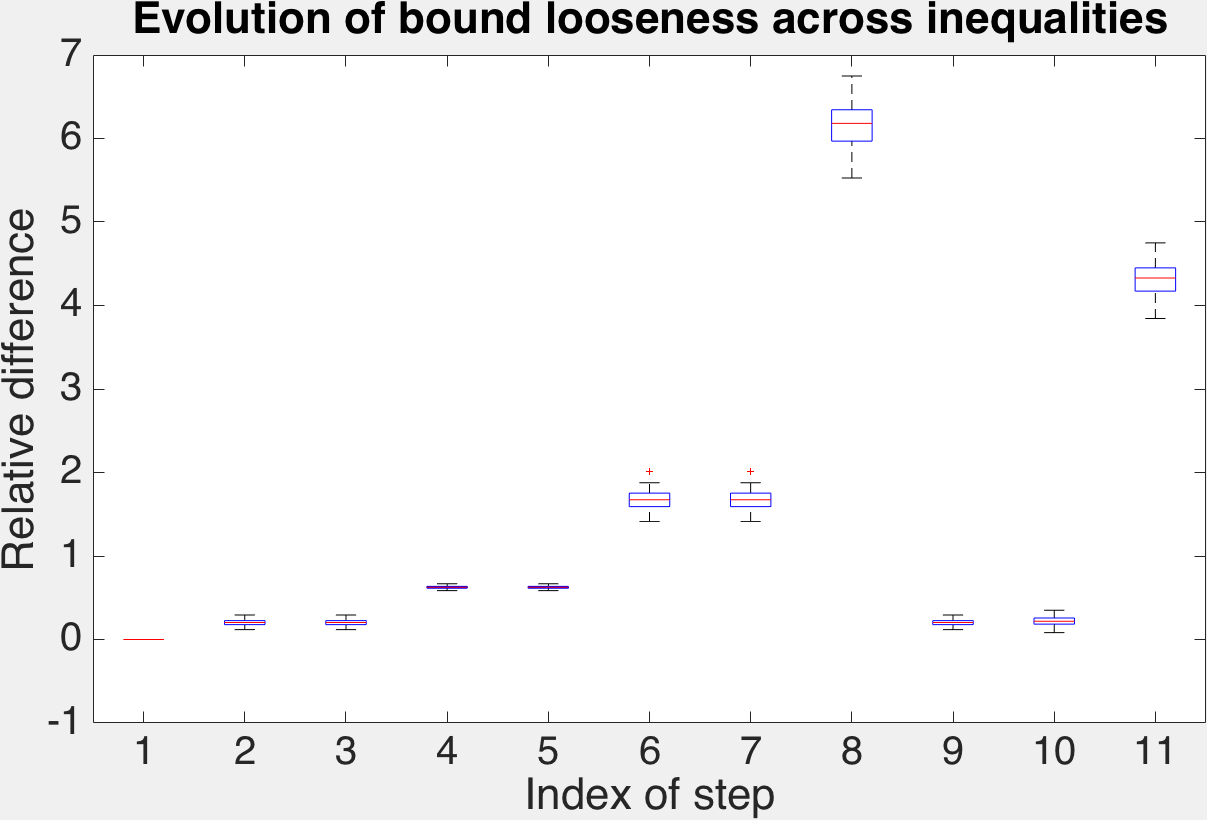}
\caption{Relative difference between inequalities in the error bound in Eq.~\ref{eq:studerBoundMain} for a $499 \times 500$ matrix drawn from a standard normal distribution. Refer to Subsection~\ref{subsec:steps} for index of step descriptions.}
\label{fig:genBoundComp1}
\end{figure}

\begin{figure}[!h]
\centering
\includegraphics[scale=0.2]{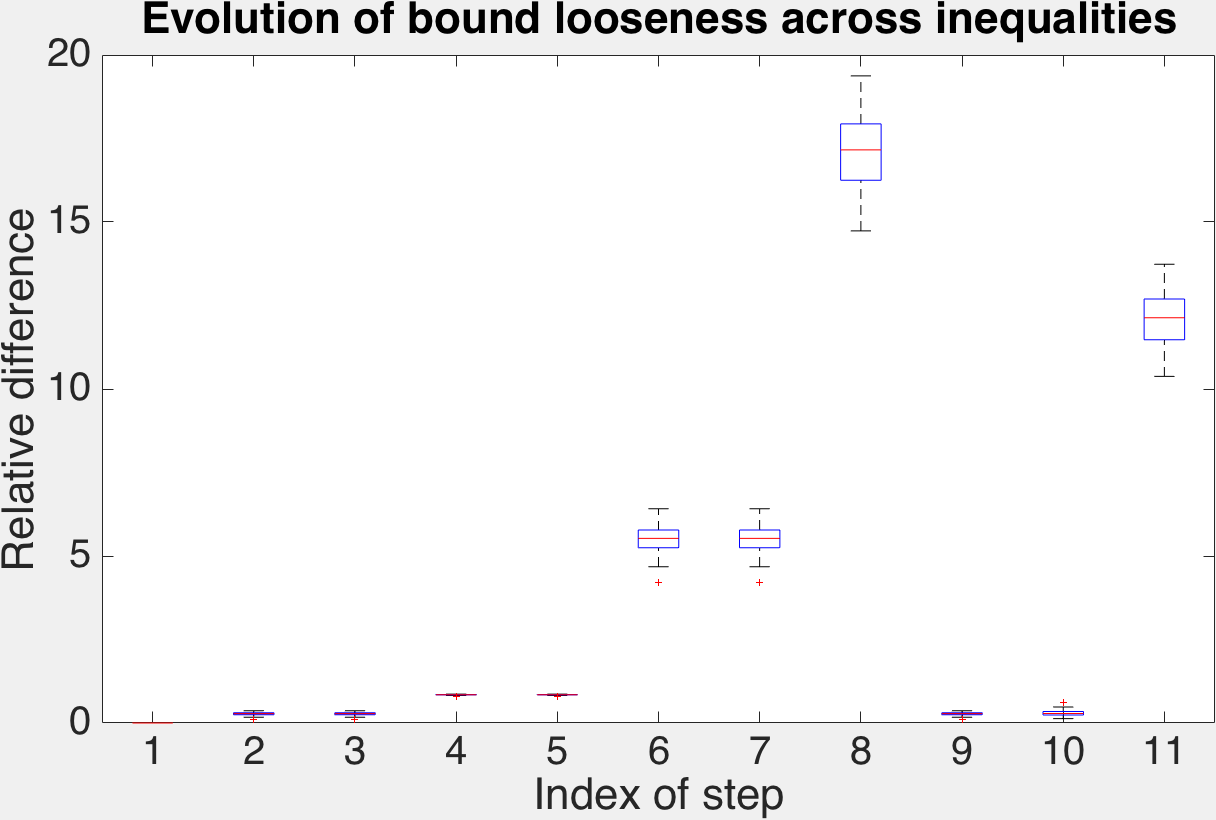}
\caption{Relative difference between inequalities in the error bound in Eq.~\ref{eq:studerBoundMain} for a $250 \times 500$ matrix drawn from a standard normal distribution. Refer to Subsection~\ref{subsec:steps} for index of step descriptions.}
\label{fig:genBoundComp2}
\end{figure}

\begin{figure}[!h]
\centering
\includegraphics[scale=0.2]{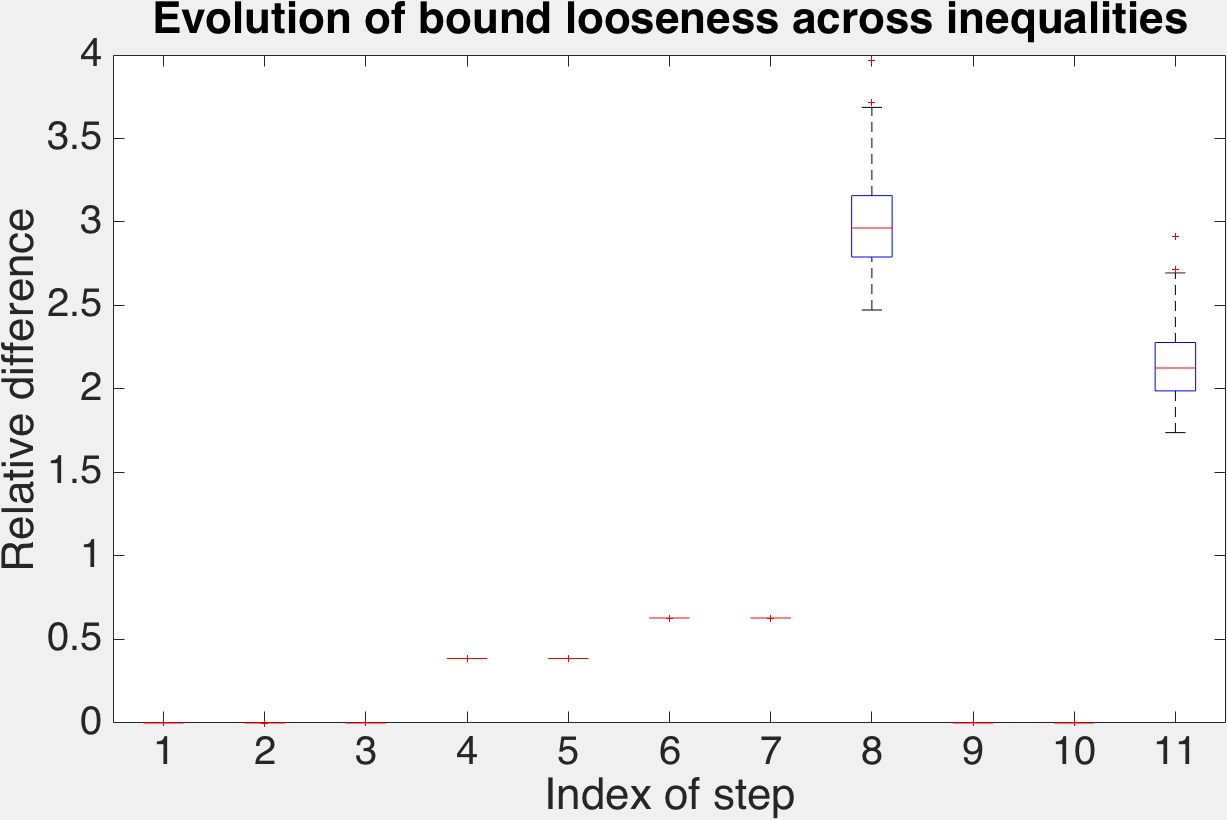}
\caption{Relative difference between inequalities in the error bound in Eq.~\ref{eq:studerBoundMain} for a $125 \times 500$ matrix drawn from a standard normal distribution. Refer to Subsection~\ref{subsec:steps} for index of step descriptions.}
\label{fig:genBoundComp4}
\end{figure}

Figs. \ref{fig:genBoundComp1}, \ref{fig:genBoundComp2} and \ref{fig:genBoundComp4} show the relative differences for these scenarios. As is immediately noticed, the differences in the first two cases are high: the bound is off by two orders of magnitude compared to the actual error surface. The other three are off by a relative error well above 1.  

Consistent among all these figures is, however, the presence of jumps between relative differences across steps. There appear jumps in the transition from step \ref{chk:step3} to step \ref{chk:step4}, and from step \ref{chk:step7} to step \ref{chk:step8} (the baseline for comparison changes between step \ref{chk:step5} and step \ref{chk:step6}, so the jump here isn't significant. Also, these two steps are rearrangements of each other, so there's no loss happening in between).

We will contrast this behavior to what happens with matrices of the kind we encounter in practical signal processing scenarios further.

% \begin{figure}[!h]
% \centering
% \includegraphics[scale=0.2]{pics/proof/736864_0582-167x500-100-randn-mat-vec-fin}
% \caption{Relative difference between inequalities in the error bound in Eq.~\ref{eq:studerBoundMain} for a $167 \times 500$ matrix drawn from a standard normal distribution. Refer to Subsection~\ref{subsec:steps} for index of step descriptions.}
% \label{fig:genBoundComp3}
% \end{figure}

% \begin{figure}[!h]
% \centering
% \includegraphics[scale=0.2]{pics/proof/736864_0586-50x500-100-randn-mat-vec-fin}
% \caption{Relative difference between inequalities in the error bound in Eq.~\ref{eq:studerBoundMain} for a $50 \times 500$ matrix drawn from a standard normal distribution. Refer to Subsection~\ref{subsec:steps} for index of step descriptions.}
% \label{fig:genBoundComp5}
% \end{figure}

\subsubsection{In the CACTI camera} \label{subsubsec:cactiLooseness}
\begin{figure}[!h]
\centering
\includegraphics[scale=0.2]{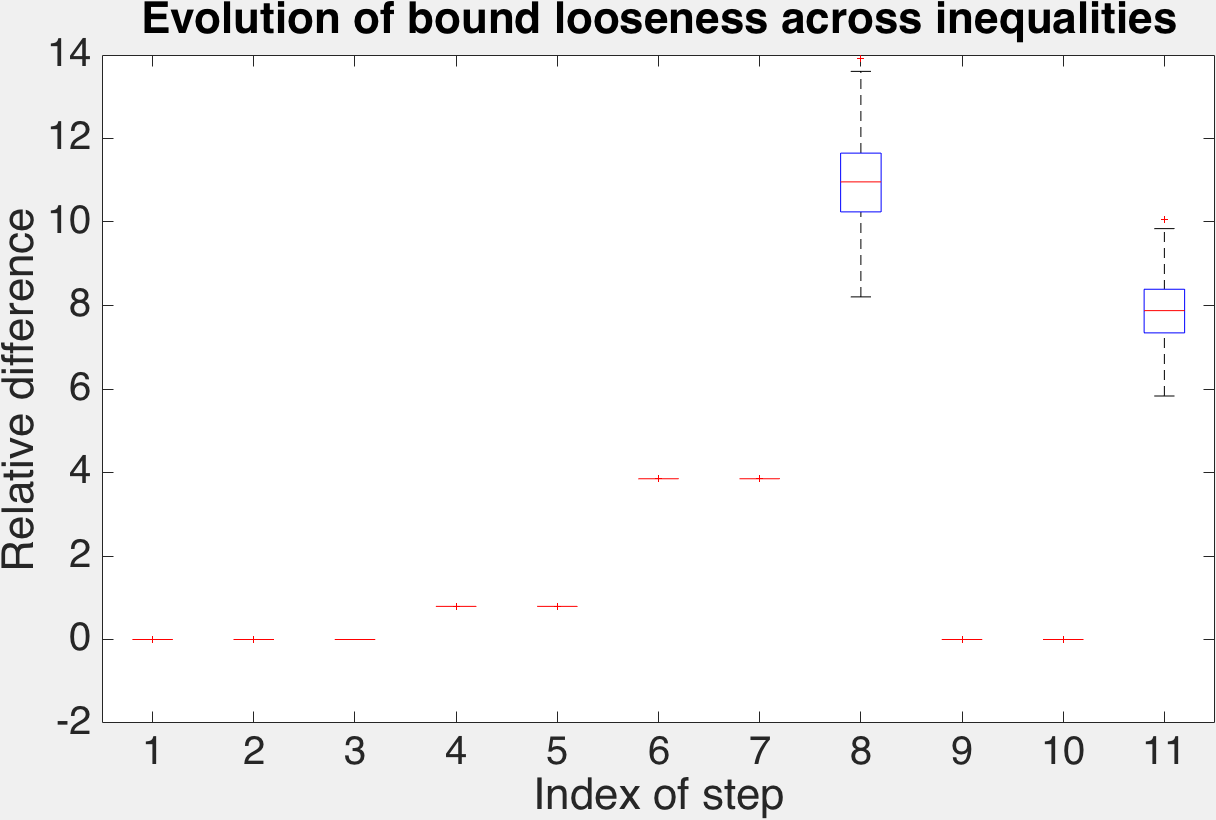}
\caption{Relative difference between inequalities in the error bound in Eq.~\ref{eq:studerBoundMain} for $8 \times 8$ random positive codes in the CACTI camera for $T = 2$. Permutations: [3, 1; 5, 5]. Refer to Subsection~\ref{subsec:steps} for index of step descriptions.}
\label{fig:cactiBoundComp2}
\end{figure}

\begin{figure}[!h]
\centering
\includegraphics[scale=0.2]{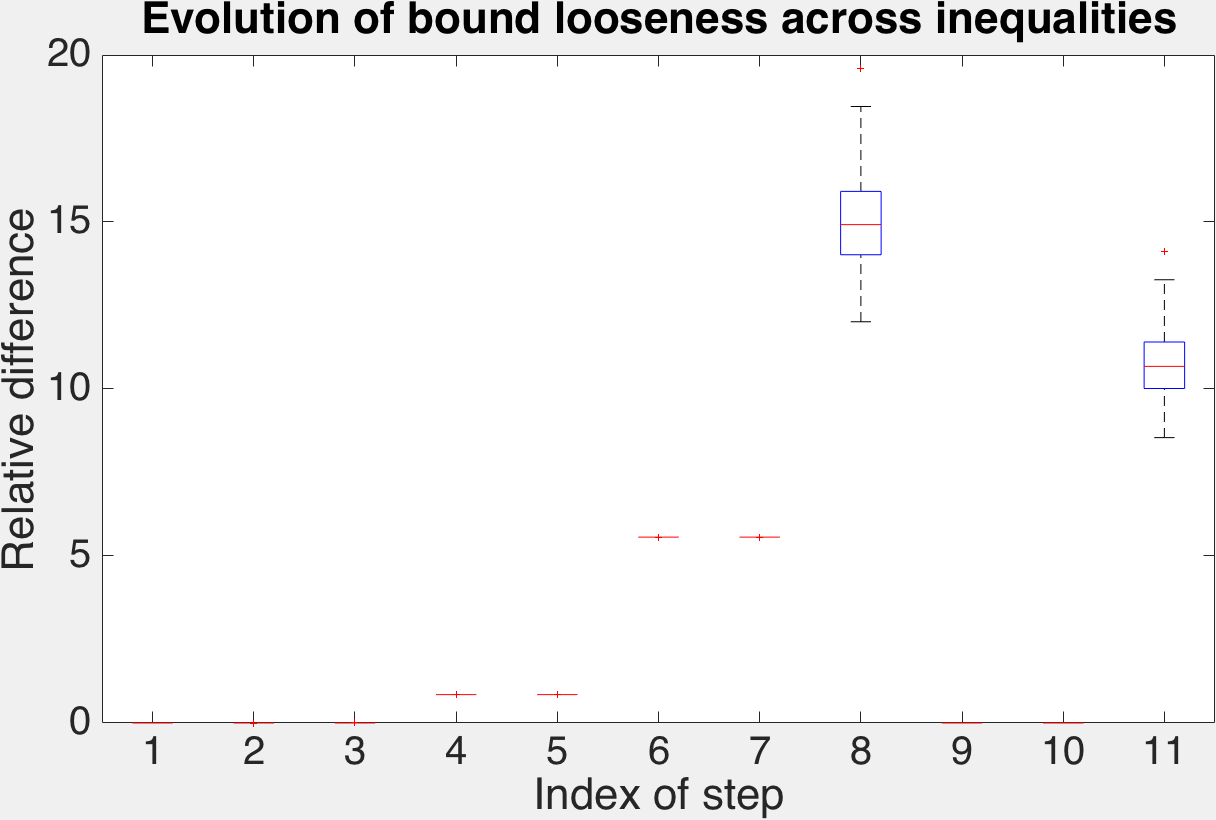}
\caption{Relative difference between inequalities in the error bound in Eq.~\ref{eq:studerBoundMain} for $8 \times 8$ random positive codes in the CACTI camera for $T = 3$. Permutations: [7, 7; 4, 4; 7, 3]. Refer to Subsection~\ref{subsec:steps} for index of step descriptions.}
\label{fig:cactiBoundComp3}
\end{figure}

\begin{figure}[!h]
\centering
\includegraphics[scale=0.2]{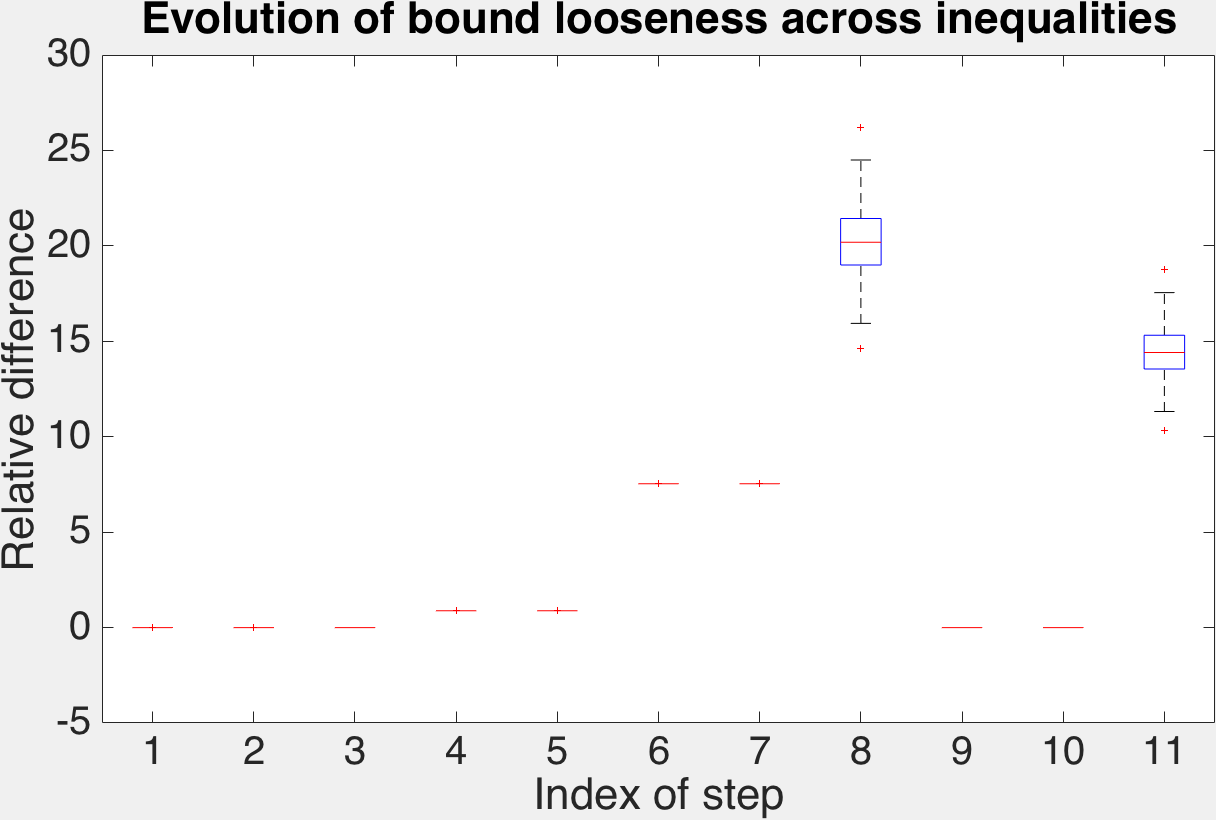}
\caption{Relative difference between inequalities in the error bound in Eq.~\ref{eq:studerBoundMain} for $8 \times 8$ random positive codes in the CACTI camera for $T = 4$. Permutations: [1, 2; 6, 6; 7, 3; 2, 7]. Refer to Subsection~\ref{subsec:steps} for index of step descriptions.}
\label{fig:cactiBoundComp4}
\end{figure}

\begin{figure}[!h]
\centering
\includegraphics[scale=0.2]{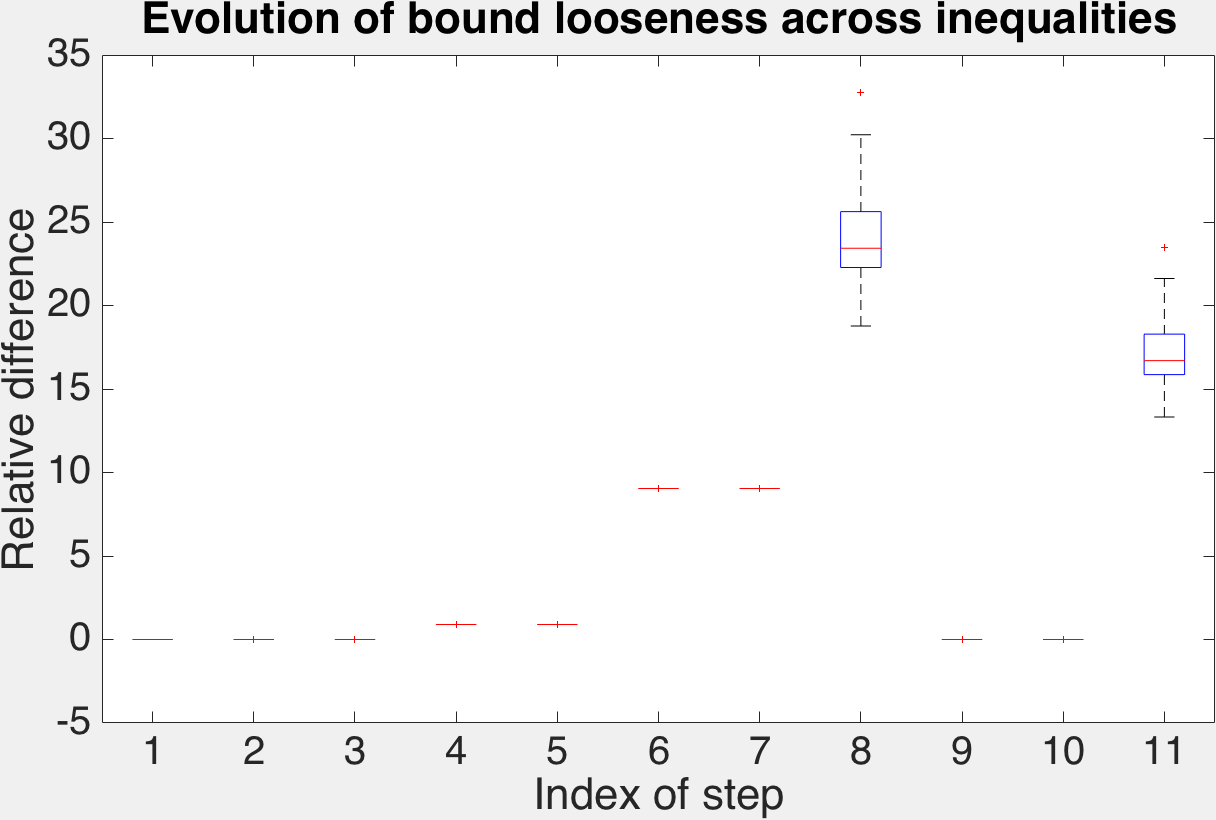}
\caption{Relative difference between inequalities in the error bound in Eq.~\ref{eq:studerBoundMain} for $8 \times 8$ random positive codes in the CACTI camera for $T = 5$. Permutations: [7, 6; 4, 8; 4, 2; 6, 5; 2, 3]. Refer to Subsection~\ref{subsec:steps} for index of step descriptions.}
\label{fig:cactiBoundComp5}
\end{figure}

\begin{figure}[!h]
\centering
\includegraphics[scale=0.2]{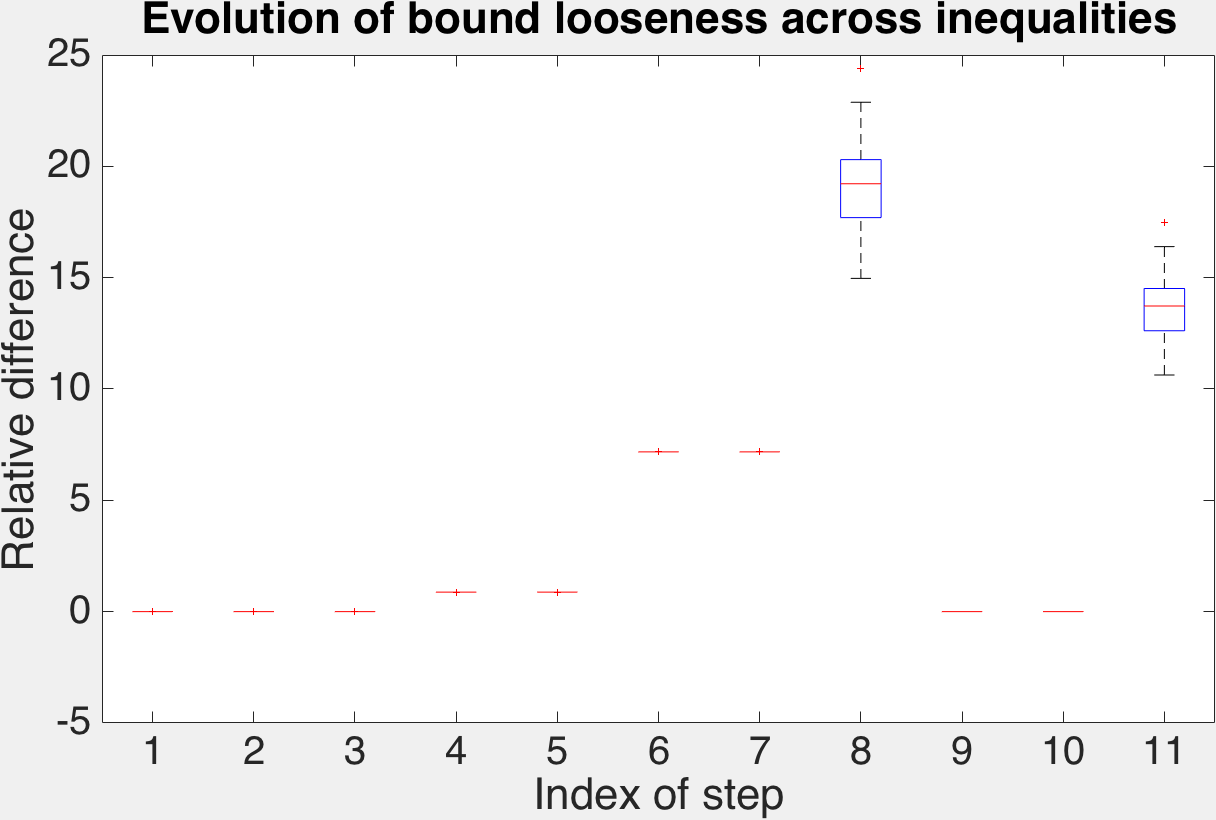}
\caption{Relative difference between inequalities in the error bound in Eq.~\ref{eq:studerBoundMain} for $8 \times 8$ random positive codes in the CACTI camera for $T = 6$. Permutations: [1, 8; 8, 7; 6, 1; 4, 6; 7, 8; 5, 3]. Refer to Subsection~\ref{subsec:steps} for index of step descriptions.}
\label{fig:cactiBoundComp6}
\end{figure}

\begin{figure}[!h]
\centering
\includegraphics[scale=0.2]{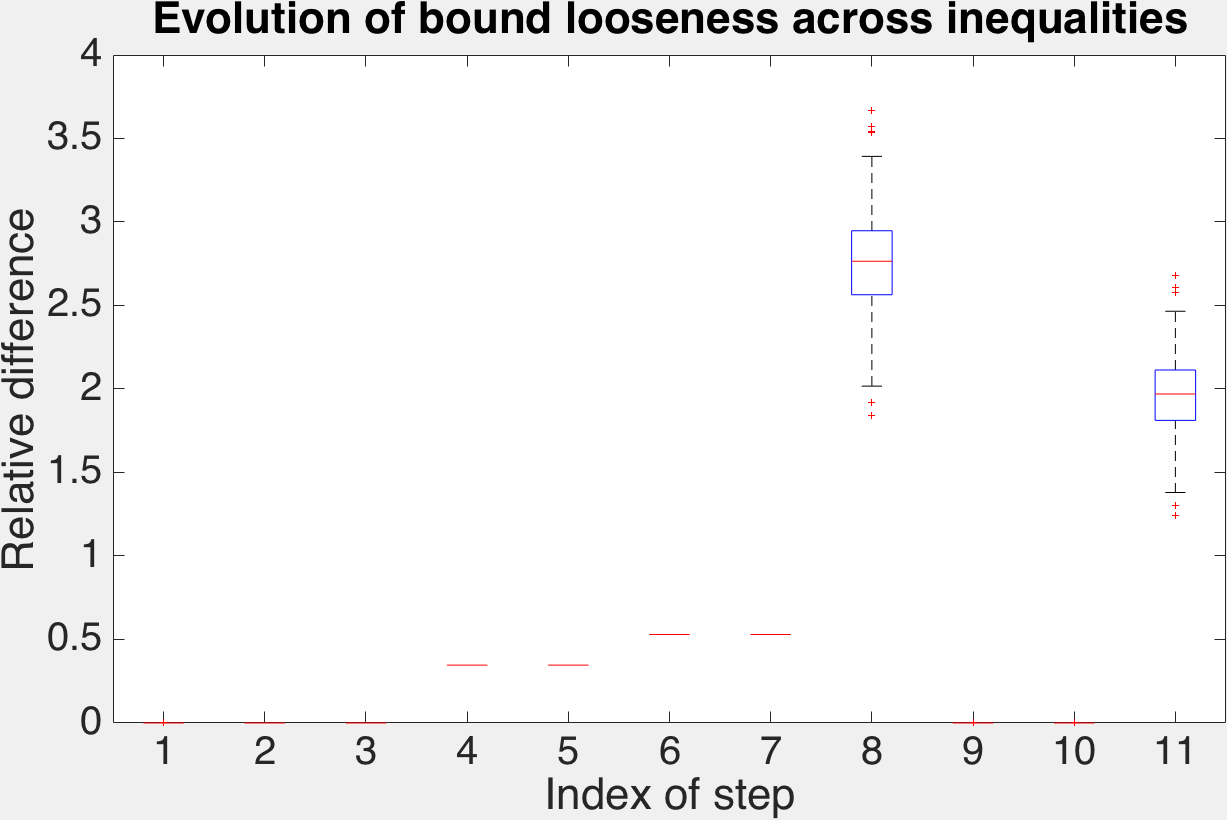}
\caption{Relative difference between inequalities in the error bound in Eq.~\ref{eq:studerBoundMain} for $8 \times 8$ designed positive codes in the CACTI camera for $T = 2$. Permutations: [5, 3; 6, 8]. Refer to Subsection~\ref{subsec:steps} for index of step descriptions.}
\label{fig:cactiDesBoundComp2}
\end{figure}

\begin{figure}[!h]
\centering
\includegraphics[scale=0.2]{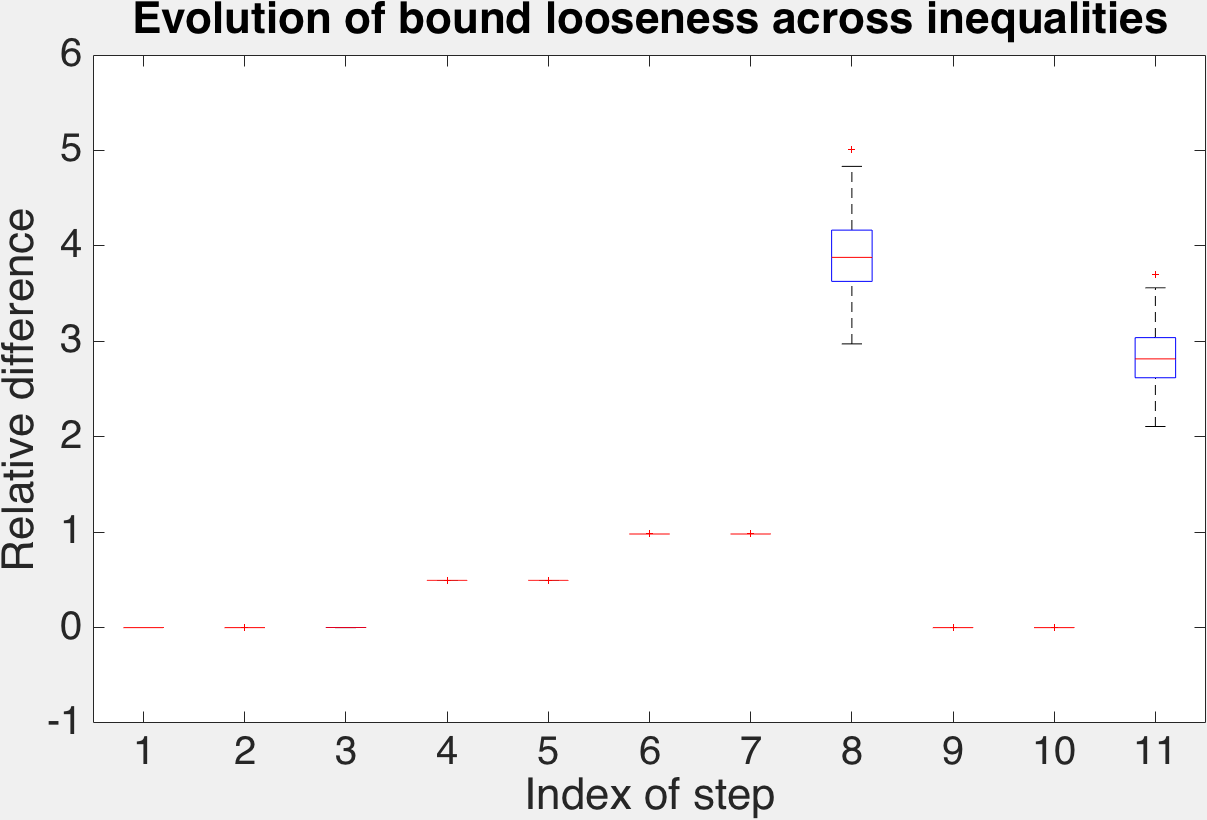}
\caption{Relative difference between inequalities in the error bound in Eq.~\ref{eq:studerBoundMain} for $8 \times 8$ designed positive codes in the CACTI camera for $T = 4$. Permutations: [7, 8; 2, 8; 6, 1; 3, 5]. Refer to Subsection~\ref{subsec:steps} for index of step descriptions.}
\label{fig:cactiDesBoundComp4}
\end{figure}

\begin{figure}[!h]
\centering
\includegraphics[scale=0.2]{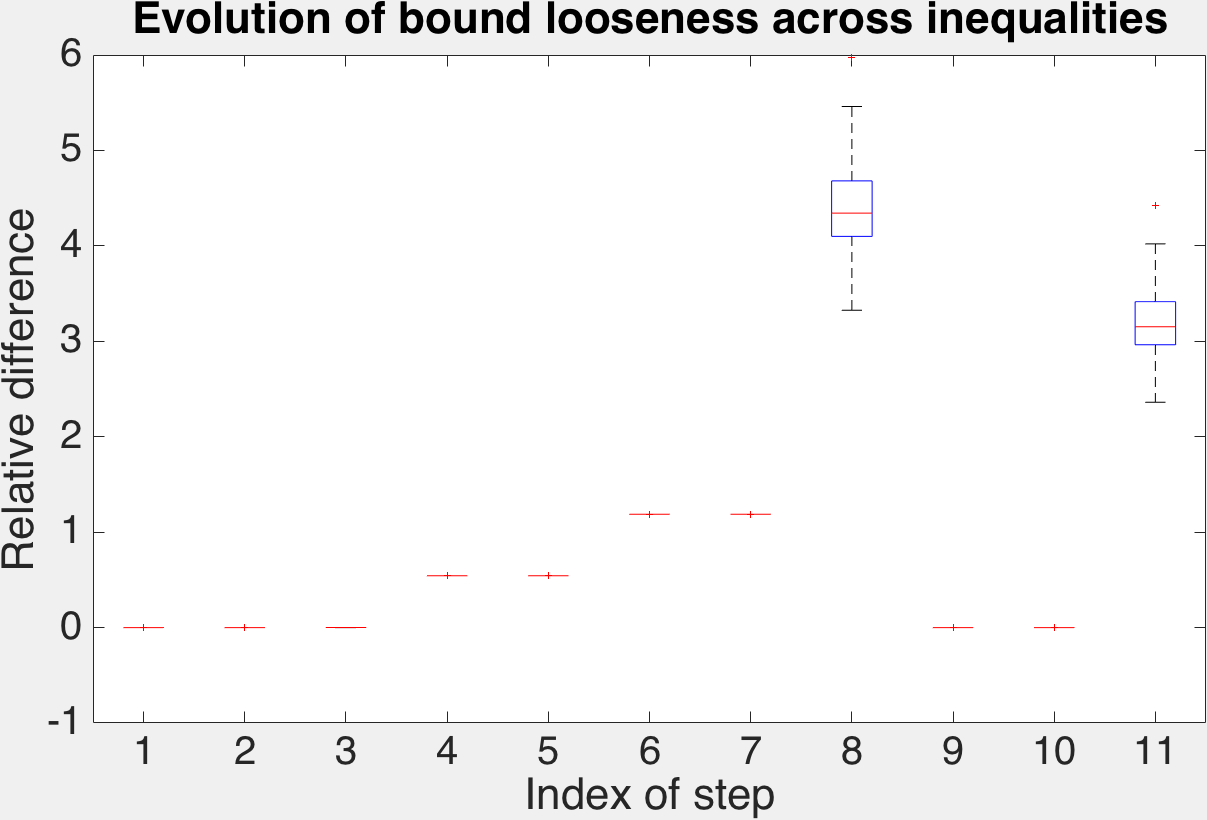}
\caption{Relative difference between inequalities in the error bound in Eq.~\ref{eq:studerBoundMain} for $8 \times 8$ designed positive codes in the CACTI camera for $T = 6$. Permutations: [6, 7; 3, 6; 6, 2; 1, 4; 8, 3; 5, 2]. Refer to Subsection~\ref{subsec:steps} for index of step descriptions.}
\label{fig:cactiDesBoundComp6}
\end{figure}

\begin{figure}[!h]
\centering
\includegraphics[scale=0.2]{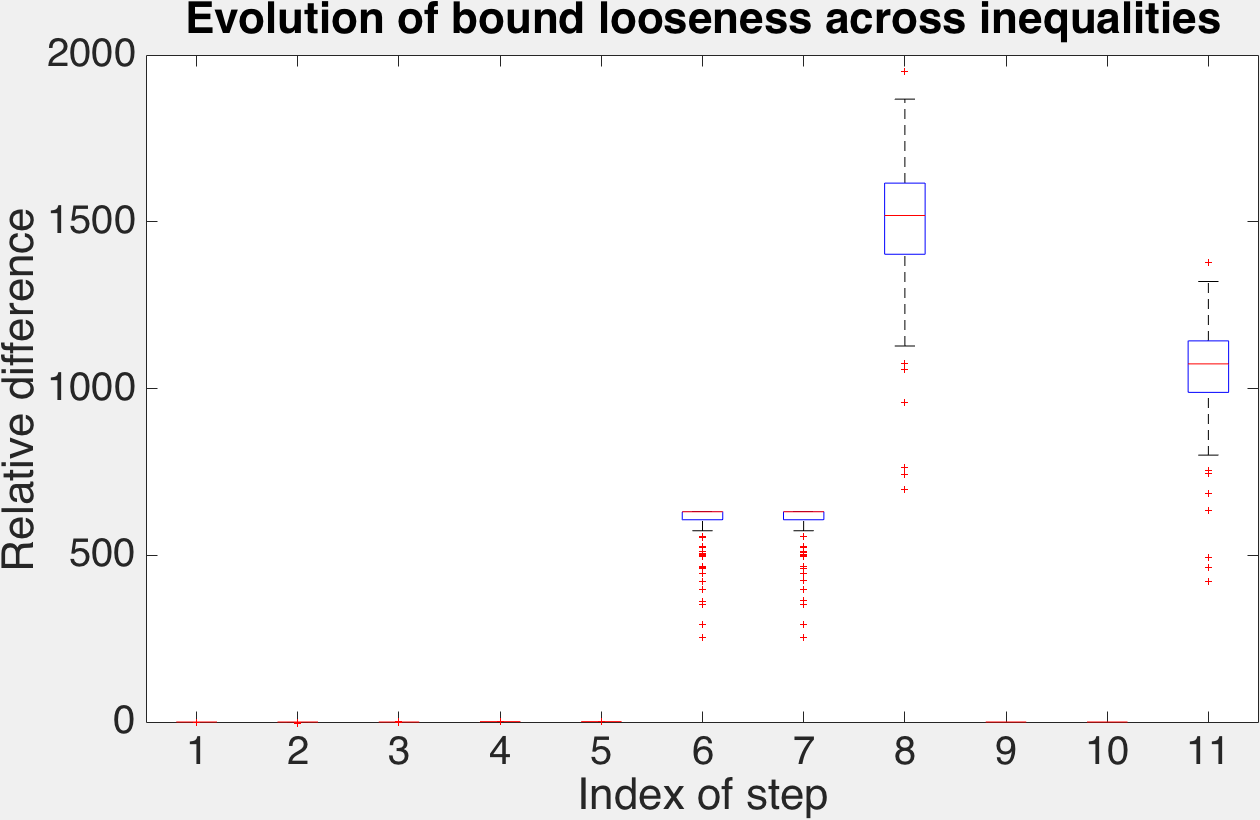}
\caption{Relative difference between inequalities in the error bound in Eq.~\ref{eq:studerBoundMain} for $8 \times 8$ average-designed positive codes in the CACTI camera for $T = 2$. Permutations: [8, 4; 7, 2]. Refer to Subsection~\ref{subsec:steps} for index of step descriptions.}
\label{fig:cactiAvgDesBoundComp2}
\end{figure}

\begin{figure}[!h]
\centering
\includegraphics[scale=0.2]{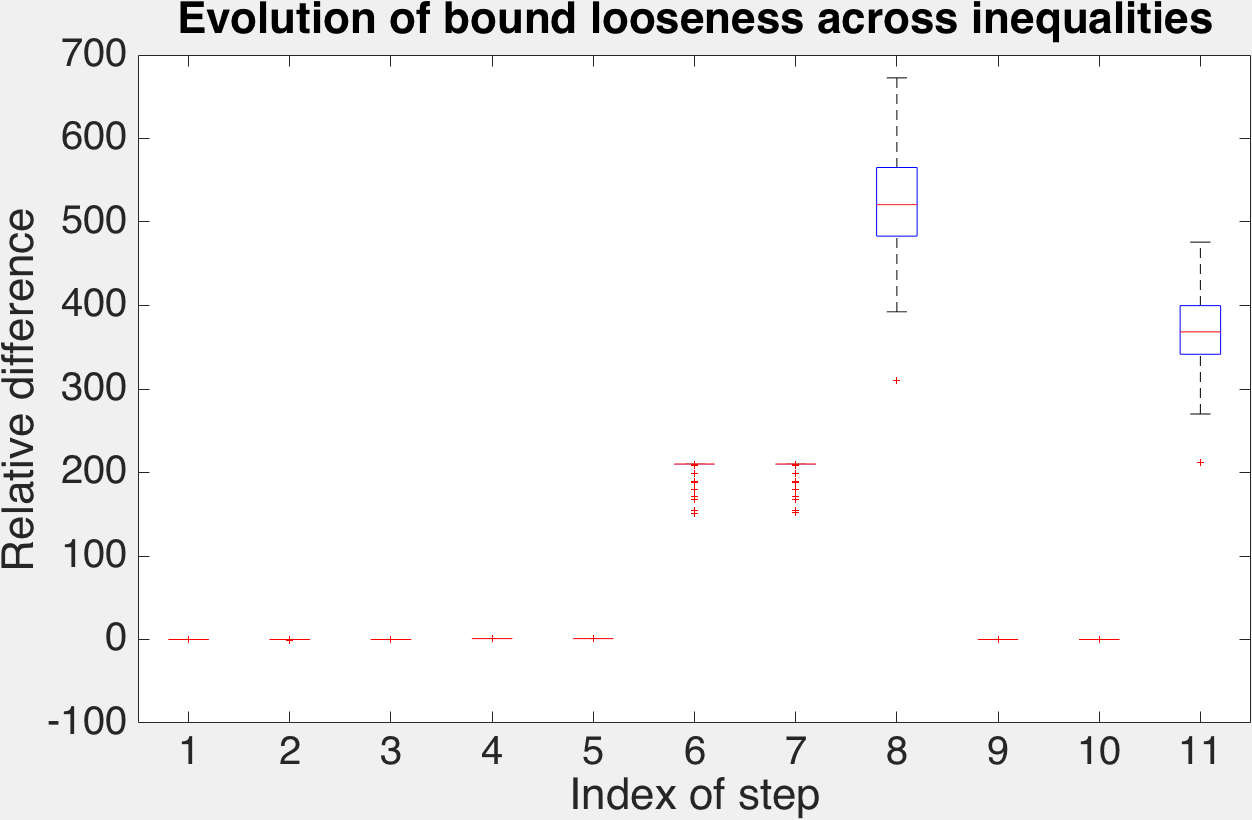}
\caption{Relative difference between inequalities in the error bound in Eq.~\ref{eq:studerBoundMain} for $8 \times 8$ average-designed positive codes in the CACTI camera for $T = 4$. Permutations: [3, 1; 1, 7; 6, 3; 8, 1]. Refer to Subsection~\ref{subsec:steps} for index of step descriptions.}
\label{fig:cactiAvgDesBoundComp4}
\end{figure}

\begin{figure}[!h]
\centering
\includegraphics[scale=0.2]{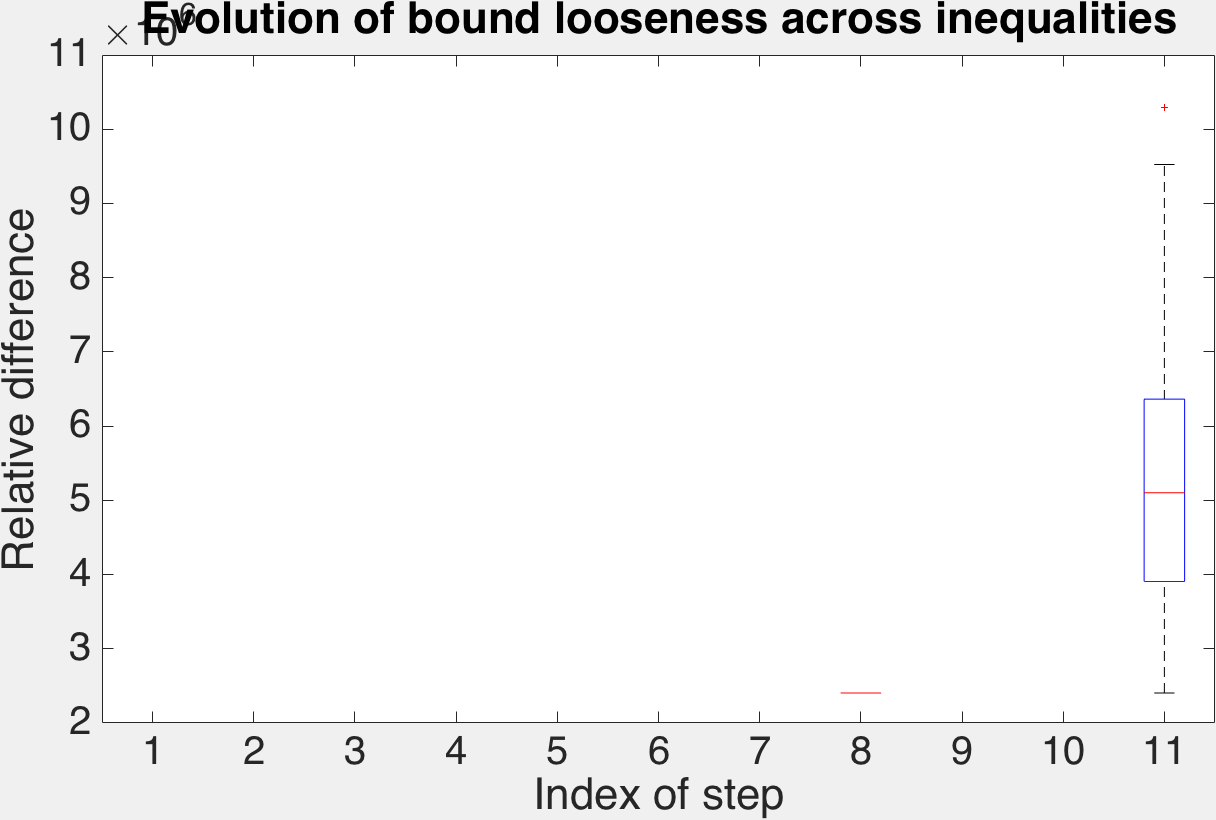}
\caption{Relative difference between inequalities in the error bound in Eq.~\ref{eq:studerBoundMain} for $8 \times 8$ average-designed positive codes in the CACTI camera for $T = 6$. Permutations: [7, 3; 5, 6; 8, 8; 5, 2; 2, 3; 7, 3]. Refer to Subsection~\ref{subsec:steps} for index of step descriptions.}
\label{fig:cactiAvgDesBoundComp6}
\end{figure}

The maximum sparsity the structure of the CACTI sensing matrices affords, in most cases, is 1, because of the $n \times nT$ size of the matrix. Typical matrices where the mask values are drawn from positive uniform [0, 1] distributions have coherence values around 0.8 to 0.9. Nevertheless, we repeat the same process as above, for $T$ = 2, 3, 4, 5 and 6 (Figs.~\ref{fig:cactiBoundComp2}, \ref{fig:cactiBoundComp3}, \ref{fig:cactiBoundComp4}, \ref{fig:cactiBoundComp5} and \ref{fig:cactiBoundComp6}), expecting the bound to work a little better in light of this low sparsity. Similarly, we also test matrices, for $T$ = 2, 4 and 6 designed for the CACTI camera in subsection \ref{subsec:cactiOpt} using both the coherence (Figs.~\ref{fig:cactiDesBoundComp2}, \ref{fig:cactiDesBoundComp4} and \ref{fig:cactiDesBoundComp6}) and the sum of squares of off-diagonal dot products of columns of the effective dictionary (Figs.~\ref{fig:cactiAvgDesBoundComp2}, \ref{fig:cactiAvgDesBoundComp4} and \ref{fig:cactiAvgDesBoundComp6}). 

It can be noticed that the same steps show significant jumps just as in the case general sensing matrices. The bound again offers enough room for failure.

\subsection{Discussion} \label{subsec:discussion}
While the figures above do not explain why coherence succeeds in traditional compressed sensing methods but doesn't in the CACTI case, the fact that there are common steps that cause significant looseness of bound is a big takeaway. These steps can be isolated and the precise inequalities causing problems can be pointed to. In our case, these are the ones leading to step \ref{chk:step4} from step \ref{chk:step3}, and from step \ref{chk:step7} to step \ref{chk:step8}. We will examine these inequalities in some detail now.

The transition from step \ref{chk:step3} to step \ref{chk:step4} involves the use of the constraint in Eq. \ref{eq:coneconstraint}. Further, going from step \ref{chk:step7} to step \ref{chk:step8} involves the constraint in Eq. \ref{eq:tubeconstraint}, and the right side of the Ger\v{s}gorin inequality in Eq. \ref{eq:gersgorindiscthm}. The culprits here, therefore, are Eq. \ref{eq:tubeconstraint}, Eq. \ref{eq:coneconstraint} and the right side of Eq. \ref{eq:gersgorindiscthm}. 

However, these are fundamental constraints that come from the problem itself. For instance, the constraint in Eq. \ref{eq:tubeconstraint} comes from the nature of the noise. The constraint in Eq.~\ref{eq:coneconstraint} comes from the fact that we use basis pursuit recovery. However, as we saw empirically in our situation, there are hardly any (in our dataset, none) vectors that meet this bound. It can be inferred, therefore, that at the cost of accommodating a set of rare worst case vectors, optimizing worst case bounds does not do well on the average case signal.

At this point, it is worthwhile to stop to consider the array of design schemes based on coherence. These methods (\cite{Elad200610, Duarte200907, Mordechay2014, Obermeier17, Bouchhima15, Abolghasemi10, Pereira14, Parada17}) attempt to design matrices with an extremely loose bound. The sheer number of these methods very emphatically states the popularity of the coherence as a measure of matrix goodness. However the analysis in our paper suggests that for very fundamental reasons, one needs to be very careful when designing sensing matrices based on coherence, despite earlier instances of success. Better, tighter bounds on sparse recovery will possibly make these methods more effective at optimizing matrices in their particular applications.

\section{A new bound} \label{sec:newBound}
\cite{Tang2015} investigates the performance bounds on sparse recovery by bounding the reconstruction error in the $l_\infty$ norm. This bound relaxes the definition of sparsity to $s(\vecx) = \norm{\vecx}{1} / \norm{\vecx}{\infty}$. This relaxation leads to a computable sufficient condition for accurate sparse recovery.

Let the compressive measurement be of the form $\vecy = \matA \vecx + \vecn$. Define $\omega_\diamond(\matA, s)$ as follows:
\begin{equation}
    \omega_\diamond(\matA, s) = \min_{s(\vecz) \leq s} \frac{\norm{\matA\vecz}{\diamond}}{\norm{\vecz}{\infty}}
    \label{eq:omegaDefinition}
\end{equation}
then it can be shown that if $\norm{\vecn}{\diamond} \leq \epsilon$ and if $\vecx$ is $k$-sparse, then the basis pursuit solver yields an $\vecxh$ that satisfies
\begin{equation}
    \norm{\hat{x} - x}{\infty} \leq \frac{2 \epsilon}{\omega_\diamond(\matA, 2k)}
\end{equation}
The bound on the $l_2$ error given by the $l_\infty$ error gives us
\begin{equation}
    \norm{\hat{x} - x}{2} \leq \frac{2 \epsilon \sqrt{2k}}{\omega_\diamond(\matA, 2k)}
    \label{eq:linfErrorBound}
\end{equation}

%Therefore, one clear way to optimize $\matA$ is to maximize the quantity $\omega_\diamond(\matA, 2k)$. If we set $\diamond = 2$,
\cite{Tang2015} shows that the quantity $\omega_2(\matA, s)$ can be written as a minimum of $n$ convex optimization problems \cite{Tang2015}
\begin{equation}
\begin{split}
    \omega_2(\matA, s) &= \min_{i \in {1..n}} \min_{\boldsymbol{\lambda} \in \mathbb{R}^{n-1}} \norm{\boldsymbol{a_i} - \matA(:, \sim i) \boldsymbol{\lambda}}{2} \\ &\text{ subject to } \norm{\boldsymbol{\lambda}}{1} \leq s-1
    \label{eq:omegaExpr}
\end{split}
\end{equation}
where $\boldsymbol{a_i}$ is the $i^\text{th}$ column of $\matA$ and $\matA(:, \sim i)$ represents the matrix $\matA$ with the $i^\text{th}$ column removed. 
% This break-up into multiple optimization problems can be exploited for optimization. The gradient of the overall objective function is the gradient of that inner objective function whose value is the highest. Simple matrix differentiation then leads to a gradient descent scheme. 

\subsection{Is this bound feasible to optimize on?} \label{subsec:lInfFeas}
The constraint that the coherence of a matrix should be low implies that no column must be written as a multiple of the other, and the pairwise projections of columns on each other should be small. In other words, low coherence implies that no column should be written as a sparse linear combination of other columns only with $k=1$. A careful look at the objective function in Eq. \ref{eq:omegaExpr} reveals that it generalizes coherence by penalizing not only expressions of one column in terms of another, but also expressions of one column as sparse combinations of other columns, where the sparsity is encouraged by the $l_1$ constraint on the linear combination.

\cite{Tang2015} claims, therefore, qualitatively, that this consideration of multiple columns, instead of pairs of columns as in the coherence, allows for a tighter bound than the one given by coherence. It is indeed true that the bound holds for all sparsity levels, in contrast to the coherence bound which holds only if $k \leq 0.5(1 + 1/\mu)$. Given the computational tractability and this greater tightness, therefore, it is a tempting thought to use this bound for matrix optimization. 

\begin{figure}[!h]
\centering
\includegraphics[scale=0.2]{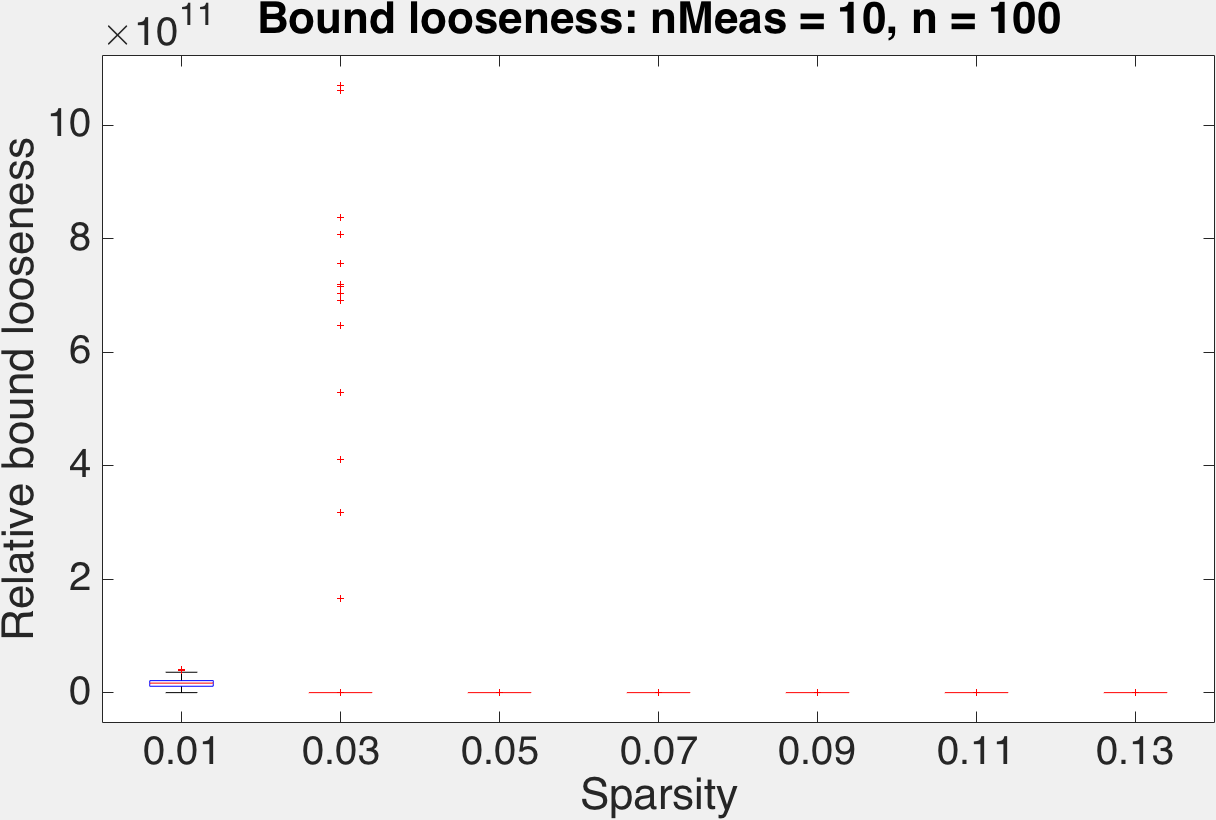}
\caption{Relative difference between reconstruction error and the error bound in Eq.~\ref{eq:linfErrorBound} as a function of sparsity for sparse $100 \times 1$ signals, sensed with a $10 \times 100$ Gaussian random matrix. Note that the median relative differences for $s > 0.01$ are still of the order of $10^6 - 10^7$, though the huge scale on the y-axis hides them.}
\label{fig:linfRRMSE1}
\end{figure}

\begin{figure}[!h]
\centering
\includegraphics[scale=0.2]{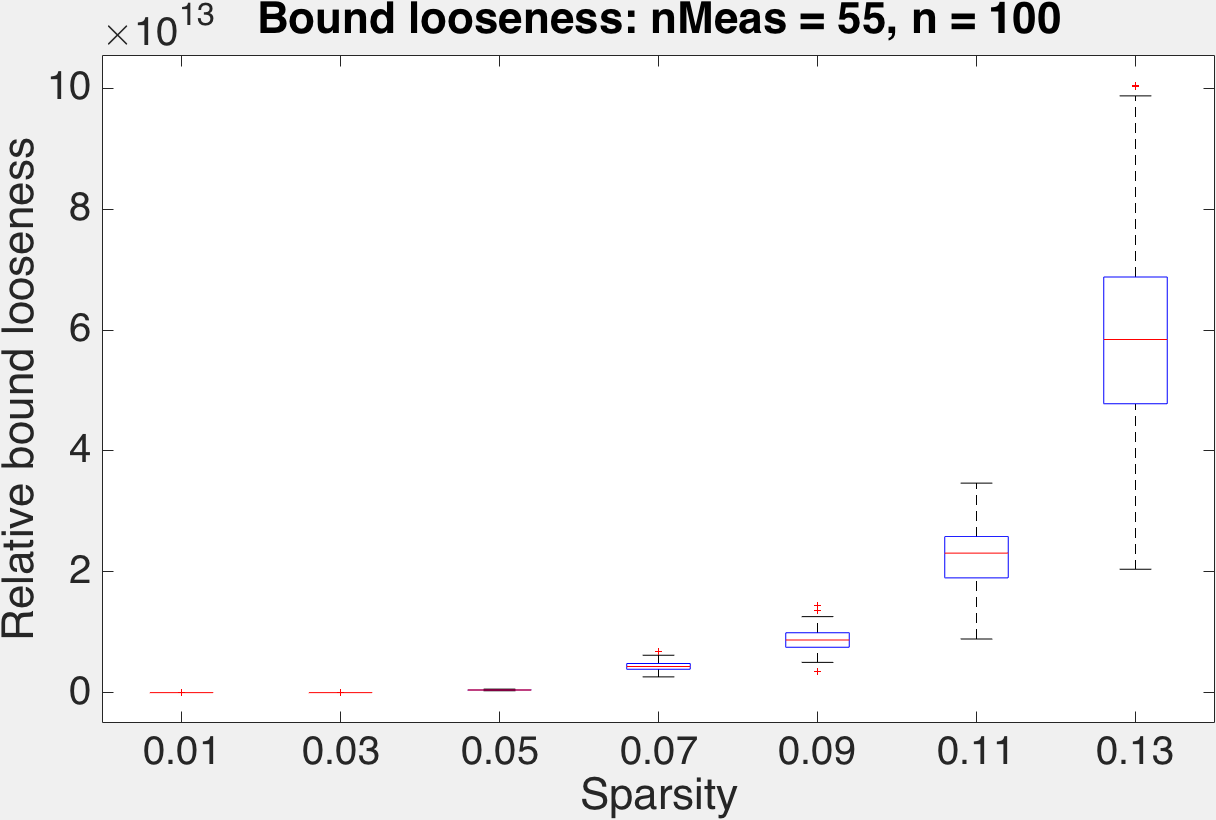}
\caption{Relative difference between reconstruction error and the error bound in Eq.~\ref{eq:linfErrorBound} as a function of sparsity for sparse $100 \times 1$ signals, sensed with a $55 \times 100$ Gaussian random matrix. Note that the median relative difference for $s = 0.01$ is still $\sim 4.82$, though the huge scale on the y-axis hides it.}
\label{fig:linfRRMSE4}
\end{figure}

\begin{figure}[!h]
\centering
\includegraphics[scale=0.2]{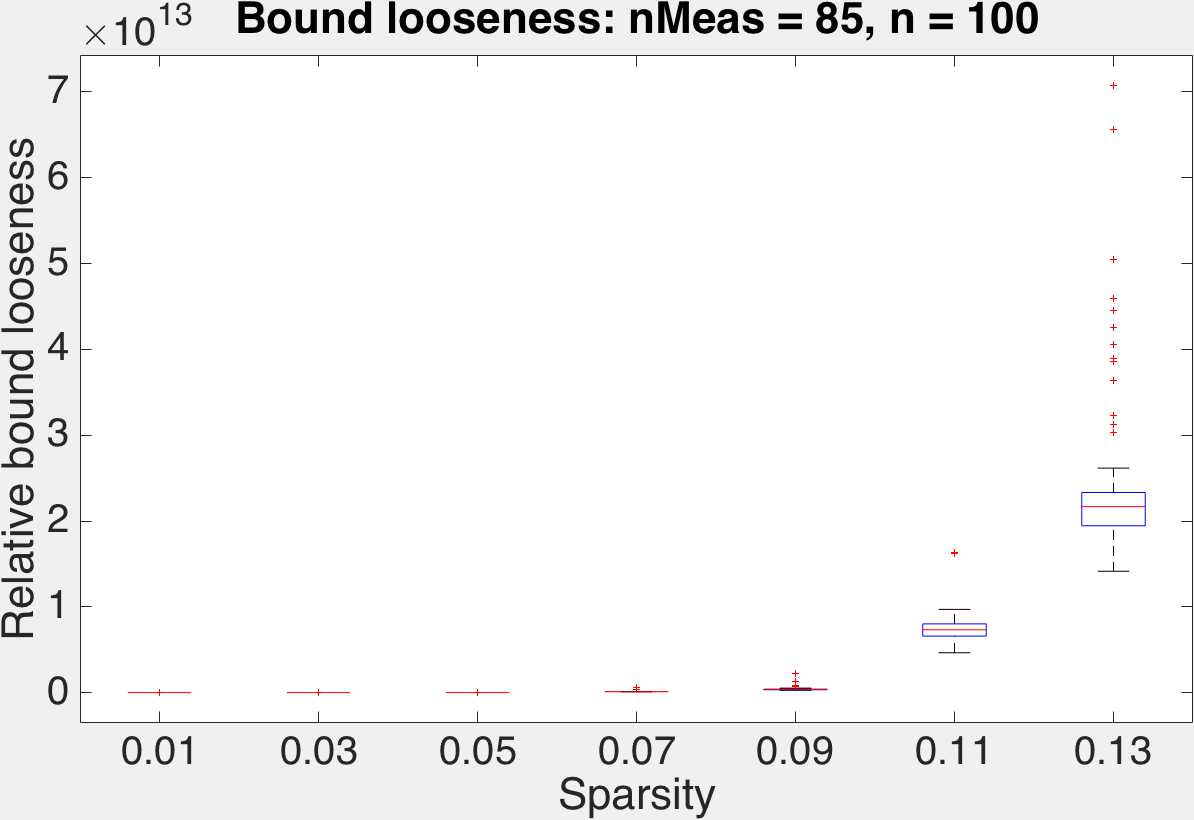}
\caption{Relative difference between reconstruction error and the error bound in Eq.~\ref{eq:linfErrorBound} as a function of sparsity for sparse $100 \times 1$ signals, sensed with a $85 \times 100$ Gaussian random matrix. Note that the median relative difference for $s = 0.01$ is still $\sim 4.51$, though the huge scale on the y-axis hides it.}
\label{fig:linfRRMSE6}
\end{figure}

Motivated by this, we run experiments to test the looseness (or tightness) of the bound in Eq. \ref{eq:linfErrorBound}. To this end, we proceed similar as we did with coherence. We generate $100 \times 1$ positive sparse vectors, varying the sparsity, and reconstruct them from noisy compressive measurements generated by a $m \times 100$ Gaussian random matrix for $m=10,\ 55,\ 85$ (Figs.~\ref{fig:linfRRMSE1}, \ref{fig:linfRRMSE4}, \ref{fig:linfRRMSE6}) and uniform noise bounded in norm by $\epsilon = 10^{-5}$, and compare the reconstruction error with the bound in Figs.~\ref{fig:genBoundComp1}, \ref{fig:genBoundComp2} and \ref{fig:genBoundComp4}.

% \begin{figure}[!h]
% \centering
% \includegraphics[scale=0.2]{pics/linf/736864_0731-100-25-100-fin}
% \caption{Relative difference between reconstruction error and the error bound in Eq.~\ref{eq:linfErrorBound} as a function of sparsity for sparse $100 \times 1$ signals, sensed with a $25 \times 100$ Gaussian random matrix. Note that the median relative difference for $s = 0.01$ is still $\sim 11.39$, though the huge scale on the y-axis hides it.}
% \label{fig:linfRRMSE2}
% \end{figure}

% \begin{figure}[!h]
% \centering
% \includegraphics[scale=0.2]{pics/linf/736864_0742-100-40-100-fin}
% \caption{Relative difference between reconstruction error and the error bound in Eq.~\ref{eq:linfErrorBound} as a function of sparsity for sparse $100 \times 1$ signals, sensed with a $40 \times 100$ Gaussian random matrix. Note that the median relative difference for $s = 0.01$ is still $\sim 5.67$, though the huge scale on the y-axis hides it.}
% \label{fig:linfRRMSE3}
% \end{figure}

% \begin{figure}[!h]
% \centering
% \includegraphics[scale=0.2]{pics/linf/736864_0771-100-70-100-fin}
% \caption{Relative difference between reconstruction error and the error bound in Eq.~\ref{eq:linfErrorBound} as a function of sparsity for sparse $100 \times 1$ signals, sensed with a $70 \times 100$ Gaussian random matrix. Note that the median relative difference for $s = 0.01$ is still $\sim 4.57$, though the huge scale on the y-axis hides it.}
% \label{fig:linfRRMSE5}
% \end{figure}

This, again, is not a very happy situation. While the bound shows promise, it works for very low sparsity levels. The coherence bound for all the Gaussian matrices we used here works for vectors that are 1-sparse, or when $s = 0.01$.

\section{Et tu, RIC?} \label{sec:ric}

The error bound introduced by \cite{Tang2015}, quoted in Eq.~\ref{eq:linfErrorBound} has been interpreted as a compromise between coherence and the RIC. Coherence penalizes the dot products of normalized column pairs, and therefore the projections of columns on each other (see Eq.~\ref{eq:cohDefn}). This qualitatively expresses the ability of one column to approximate the other, or in other words, how well a $1$-sparse combination of $n-1$ of columns from $\matA$ can represent the remaining column. Looking at Eq.~\ref{eq:omegaExpr}, one realizes that the linear combination being penalized here is not just $1$-sparse: all linear combinations of $n-1$ columns with a coefficient vector $\boldsymbol{\lambda}$ are penalized for how well they can represent the $n^\text{th}$ column, as long as $\|\boldsymbol{\lambda}\|_1 \leq s$. While no theoretical claims can be made about which bound is better, a looseness analysis of the RIC bound seems to be a fitting final section of this paper.

\begin{figure}[!h]
\centering
\includegraphics[scale=0.2]{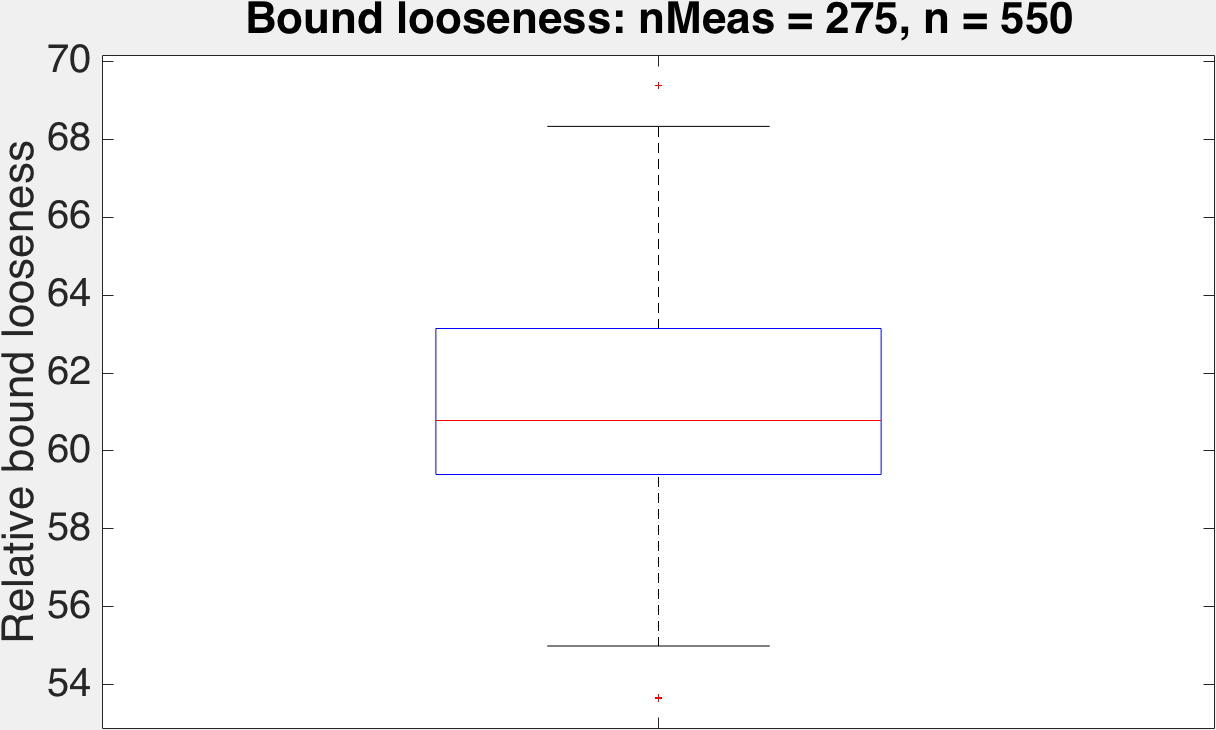}
\caption{Relative difference between reconstruction error and the error bound in Eq.~\ref{eq:ricErrorBound} for sparse $550 \times 1$ signals with $k = 2$, sensed with a $275 \times 550$ Gaussian random matrix}
\label{fig:ricRRMSE1}
\end{figure}

\begin{figure}[!h]
\centering
\includegraphics[scale=0.2]{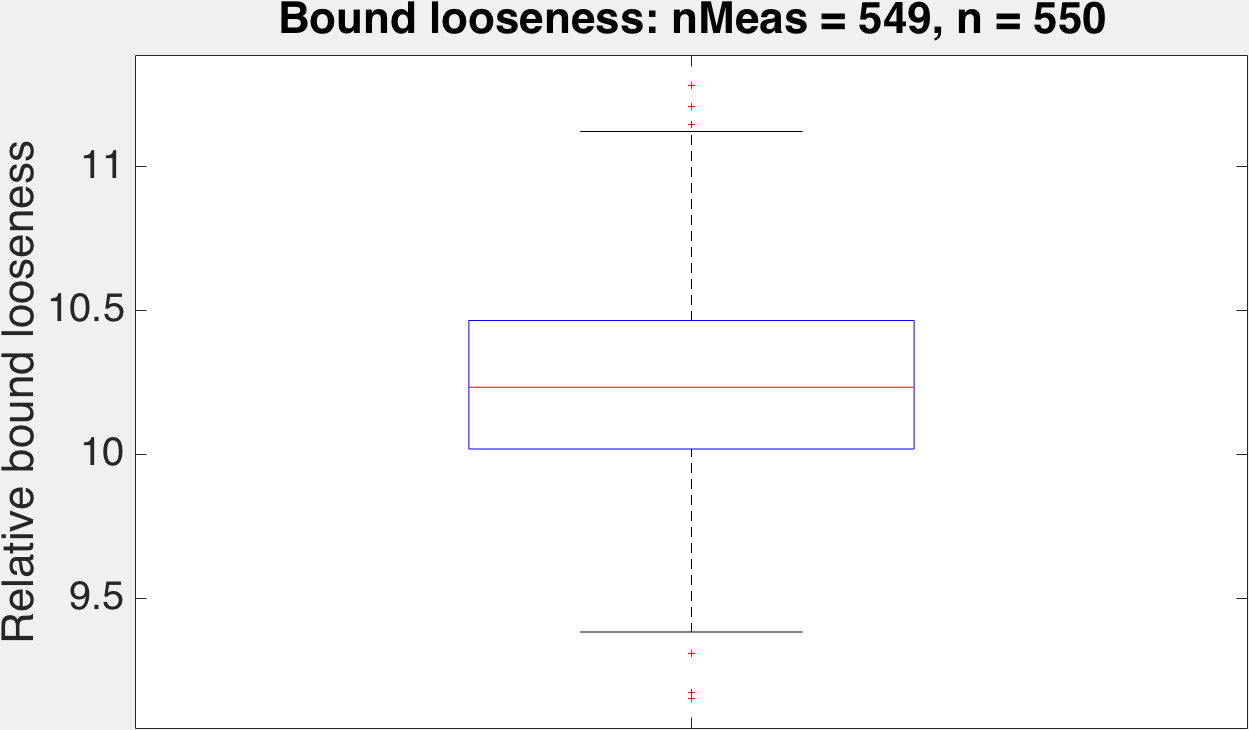}
\caption{Relative difference between reconstruction error and the error bound in Eq.~\ref{eq:ricErrorBound} sparse $550 \times 1$ signals with $k = 2$, sensed with a $549 \times 550$ Gaussian random matrix}
\label{fig:ricRRMSE4}
\end{figure}

\begin{figure}[!h]
\centering
\includegraphics[scale=0.2]{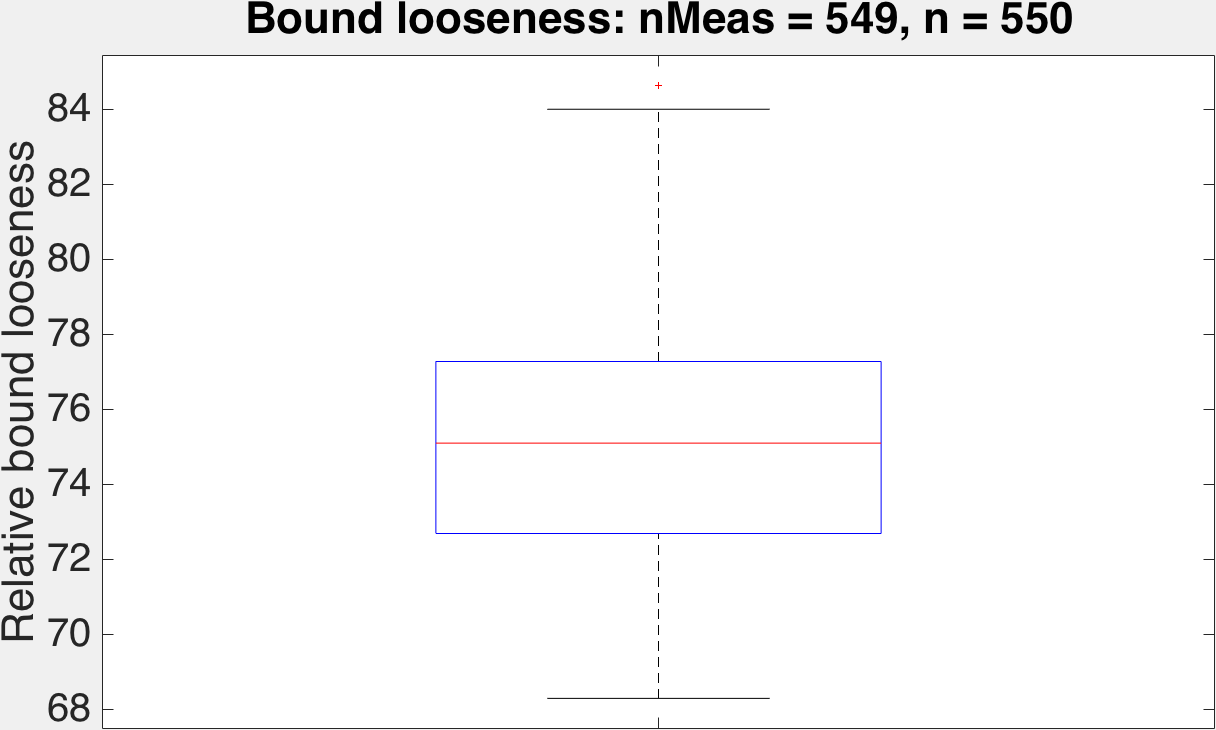}
\caption{Relative difference between reconstruction error and the error bound in Eq.~\ref{eq:ricErrorBound} for sparse $550 \times 1$ signals with $k = 3$, sensed with a $549 \times 550$ Gaussian random matrix}
\label{fig:ricRRMSE5}
\end{figure}

The RIC-based bound error bound in \cite{Cai2010} states that if $\delta_k < 0.307$
\begin{equation}
    \|\vecxh - \vecx\|_2 \leq \frac{\epsilon}{0.307-\delta_k}.
    \label{eq:ricErrorBound}
\end{equation}
Though the RIC is intractable to compute, it is computable for small sparsity levels for reasonably sized matrices. We, therefore, calculate the relative difference between the left and right hand sides of the bound in Eq.~\ref{eq:ricErrorBound} with respect to the left hand side, which is the actual error between the actual vector and the reconstruction. We randomly generate $m \times 550$ matrices for $m$ = 275 and 549 (Figs. \ref{fig:ricRRMSE1} and \ref{fig:ricRRMSE4} respectively), and random positive $k = 2$-sparse $550 \times 1$ vectors. These numbers are selected so that the RIC condition $\delta_k \leq 0.307$ is satisfied. Then, reconstructing using the basis pursuit solver in Eq.~\ref{eq:basisPursuit}, we calculate the $l_2$ error between the original and reconstructed vectors. We also perform the same analysis on a $k = 3$-sparse vector for a $549 \times 500$ matrix (Fig.~\ref{fig:ricRRMSE5}), since the matrix instance we chose permits the RIC condition to hold.

% \begin{figure}[!h]
% \centering
% \includegraphics[scale=0.2]{pics/ric/736864_0829-367x550-2-100}
% \caption{Relative difference between reconstruction error and the error bound in Eq.~\ref{eq:ricErrorBound} for sparse $550 \times 1$ signals with $k = 2$, sensed with a $367 \times 550$ Gaussian random matrix}
% \label{fig:ricRRMSE2}
% \end{figure}

% \begin{figure}[!h]
% \centering
% \includegraphics[scale=0.2]{pics/ric/736864_0846-412x550-2-100}
% \caption{Relative difference between reconstruction error and the error bound in Eq.~\ref{eq:ricErrorBound} for sparse $550 \times 1$ signals with $k = 2$, sensed with a $412 \times 550$ Gaussian random matrix}
% \label{fig:ricRRMSE3}
% \end{figure}

The values of looseness of bound in the RIC case are not close to zero either. The RIC, therefore, does not establish a tight bound on the recovery error. The problem optimizing with coherence now, we establish empirically, is twofold: the coherence establishes a loose bound on the RIC, and the RIC establishes a loose bound on the recovery error. 

\section{The average case: A proof of concept} \label{sec:mmse}
Investigating the source of looseness in compressed sensing bounds led us to a dead end: the looseness comes from steps that are at the core of the problem themselves. We saw in Subsection~\ref{subsec:discussion}, for instance, how bounding the measurement noise was a major source of error, but a major step in the proof as well. The conclusion, then, was that while the error bound takes into consideration all vectors in the input space, the fact that the error bound is not met in a large dataset of vectors we simulated makes the consideration seem unnecessary. The bound is loose at the expense of taking into account a low probability set of vectors.

We now note that any compressed sensing bound, being universal to all vectors, characterizes recovery error in terms of sensing matrix properties and signal sparsity. No other property of the input signal is used. Therefore, in general, any approach targeting worst case errors in terms of these quantities will encounter the same worst case low probability vector pitfall as coherence. A possible direction of future work, therefore, is to circumvent these vectors by considering an average case error analysis.

The average case minimum mean square error (MMSE), introduced in \cite{Carson2012} as
\begin{equation}
\mathbb{E}_{\boldsymbol{X}} \left[ tr \left\{ \left( \vecx - \mathbb{E}_{\boldsymbol{X}|\boldsymbol{Y}} [\vecx|\vecy] \right) \left( \vecx - \mathbb{E}_{\boldsymbol{X}|\boldsymbol{Y}} [\vecx|\vecy] \right)^T  \right\} \right] \label{eq:mmse} \\
\end{equation}
is successfully lower-bounded in the same paper by a function of the mutual information between $\boldsymbol{x}$ and $\boldsymbol{y}=\boldsymbol{\Phi x}$. The aim is to design $\boldsymbol{\Phi}$ so as to optimize the mutual information between $\boldsymbol{x}$ and $\boldsymbol{y}=\boldsymbol{\Phi x}$. The mutual information along with an entropy term is proved to be a lower bound, but that certainly may not necessarily decrease the average recovery error.

A full analysis of average case error, however, requires estimating a posterior on the space of input vectors as well as assuming a specific statistical model for noise, and involves the MMSE in Eq.~\ref{eq:mmse}, an intractable quantity to calculate for many commonly occurring priors and likelihoods. Such an analysis is beyond the scope of this paper. A sampling-based approach with an appropriate prior over the space of input signals seems to be a direction in which to proceed for any tractable sensing matrix design considering the average case. This method, however, will require coming up with a prior distribution such that the posterior $\mathbb{E}[\vecx|\vecy]$ can be sampled from. The parametrization can then be used to optimize sensing matrices with respect to the expectation.

For the purposes of providing a proof-of-concept, we circumvent the calculation of the MMSE as follows: we take a dataset of random vectors from the prior distribution we are interested in, use basis pursuit for signal recovery instead of the full Bayesian $\mathbb{E}[\vecx|\vecy]$. The MMSE is then approximated as the mean squared error on this dataset. With this, we perform a random search on the sensing matrix $\boldsymbol{\Phi}$. In the case of the CACTI matrices considered in this paper, we show superior recovery performance with matrices designed in this manner as compared to coherence. In the case of general sensing matrices, where the number of degrees of freedom is larger, the performance is no worse than that of coherence.

\subsection{General sensing matrices} \label{subsec:genMMSEOpt}
\begin{figure}
    \centering
    \includegraphics[scale=0.2]{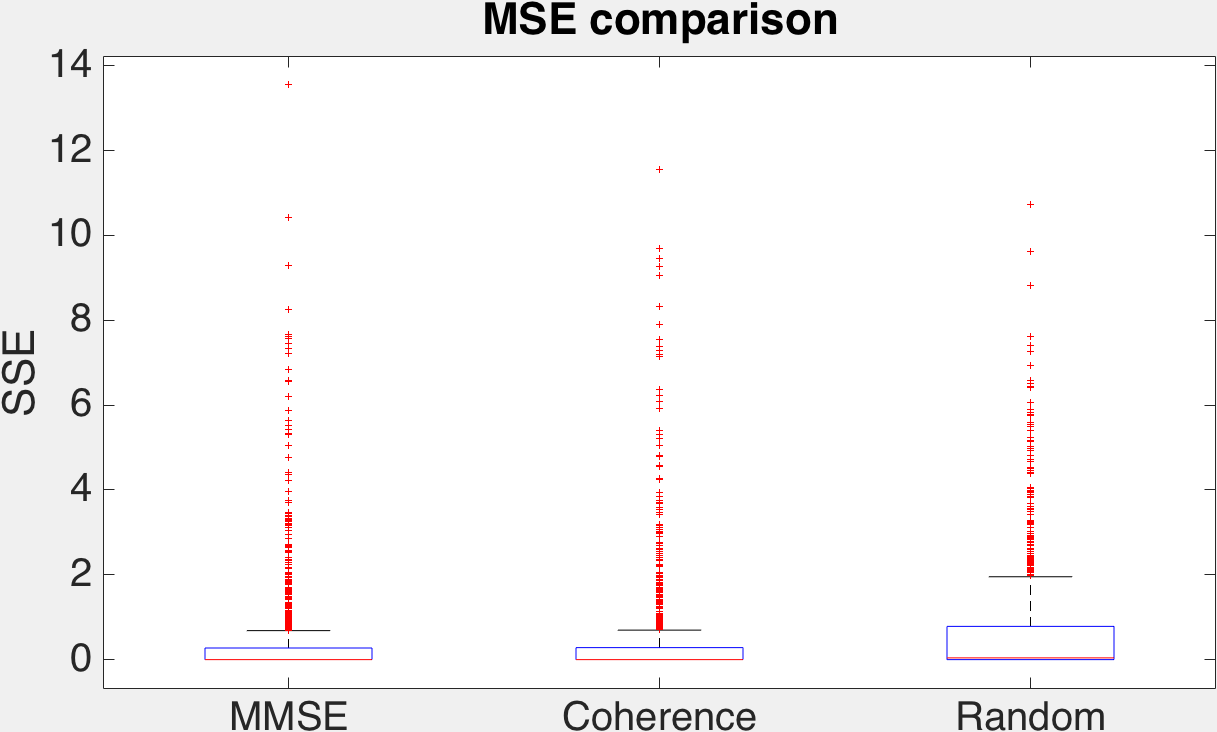}
    \caption{Comparison of sum of squared errors on a dataset of 250 sparse vectors, between $8 \times 50$ general sensing matrices optimized with mean square error, optimized with coherence and a general random matrix, targeted at sparsities of $0.08$}
    \label{fig:genMMSE2}
\end{figure}

\begin{figure}
    \centering
    \includegraphics[scale=0.2]{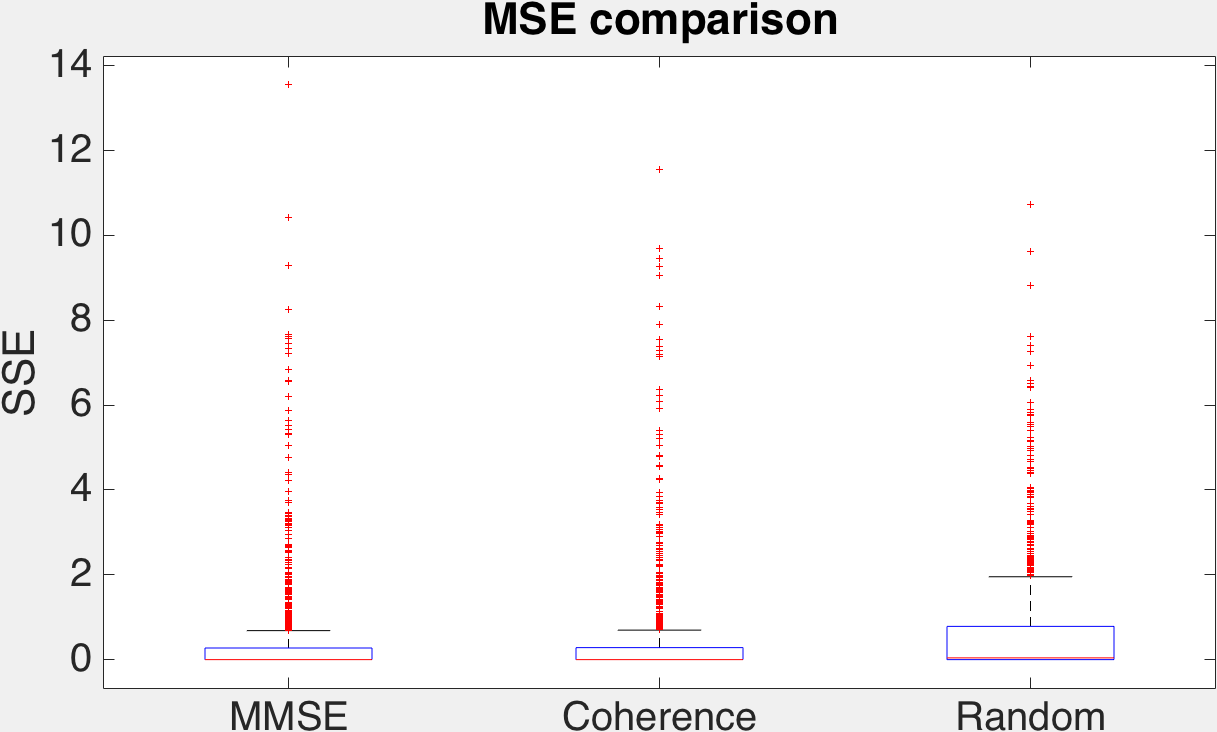}
    \caption{Comparison of sum of squared errors on a dataset of 250 sparse vectors, between $25 \times 50$ general sensing matrices optimized with mean square error, optimized with coherence and a general random matrix, targeted at sparsities of $0.2$}
    \label{fig:genMMSE6}
\end{figure}

We design $m \times 50$ matrices for $m = $ 8 and 25 (Figs.~\ref{fig:genMMSE2} and \ref{fig:genMMSE6} respectively) targeted at perfectly sparse vectors with sparsities of $s = $ 0.2 and 0.08 respectively. We take a set $\{\mathcal{S}_1, \mathcal{S}_2, \cdots \mathcal{S}_z\}$ of $z=3$ randomly generated support sets, and draw random vectors having supports from this set (this is done so that we do not have to explore the entire set of sparse vectors at a given sparsity, limiting the number of sparse recovery problems we have to solve per objective function evaluation, and hence reducing computational complexity). A random search is done on the elements of the matrix to minimize the squared error on this dataset, limiting the number of samples per iteration to 10. We then compare, on an independently generated test set of vectors having supports from these sets, the performance of a matrix designed to minimize coherence and the random matrix we started from with our matrix. It is clear from these boxplots that our method does no worse than coherence-based design in the general sensing matrix case.

% \begin{figure}
%     \centering
%     \includegraphics[scale=0.2]{pics/mmse/gen-mmse-comparison-2-support}
%     \caption{Comparison of sum of squared errors on a dataset of 250 sparse vectors, between $12 \times 50$ general sensing matrices optimized with mean square error, optimized with coherence and a general random matrix, targeted at sparsities of $0.12$}
%     \label{fig:genMMSE4}
% \end{figure}

\subsection{In the CACTI camera} \label{subsec:CACTIMMSEOpt}
\begin{figure}
    \centering
    \includegraphics[scale=0.2]{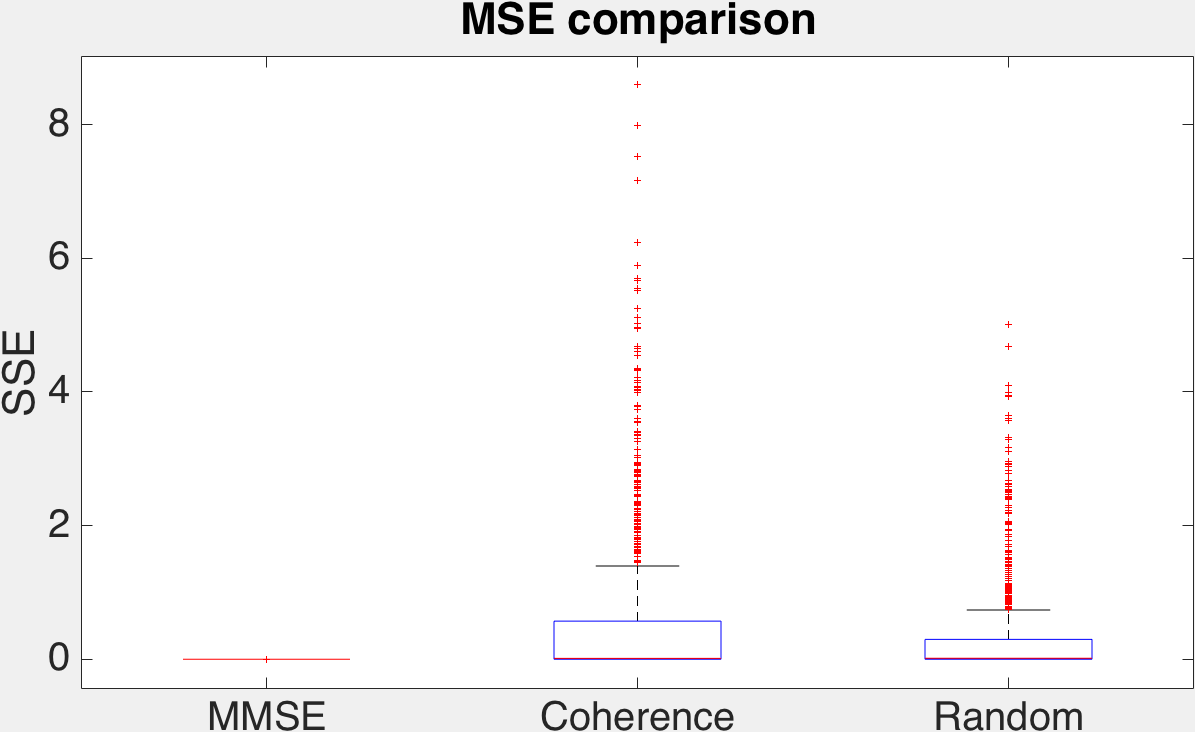}
    \caption{Comparison of sum of squared errors on a dataset of 250 sparse vectors, using $8 \times 8$ positive codes with $T=2$, optimized using the mean square error, optimized with coherence and a positive random code, targeted at sparsity of $0.2$}
    \label{fig:cactiMMSE2}
\end{figure}

\begin{figure}
    \centering
    \includegraphics[scale=0.2]{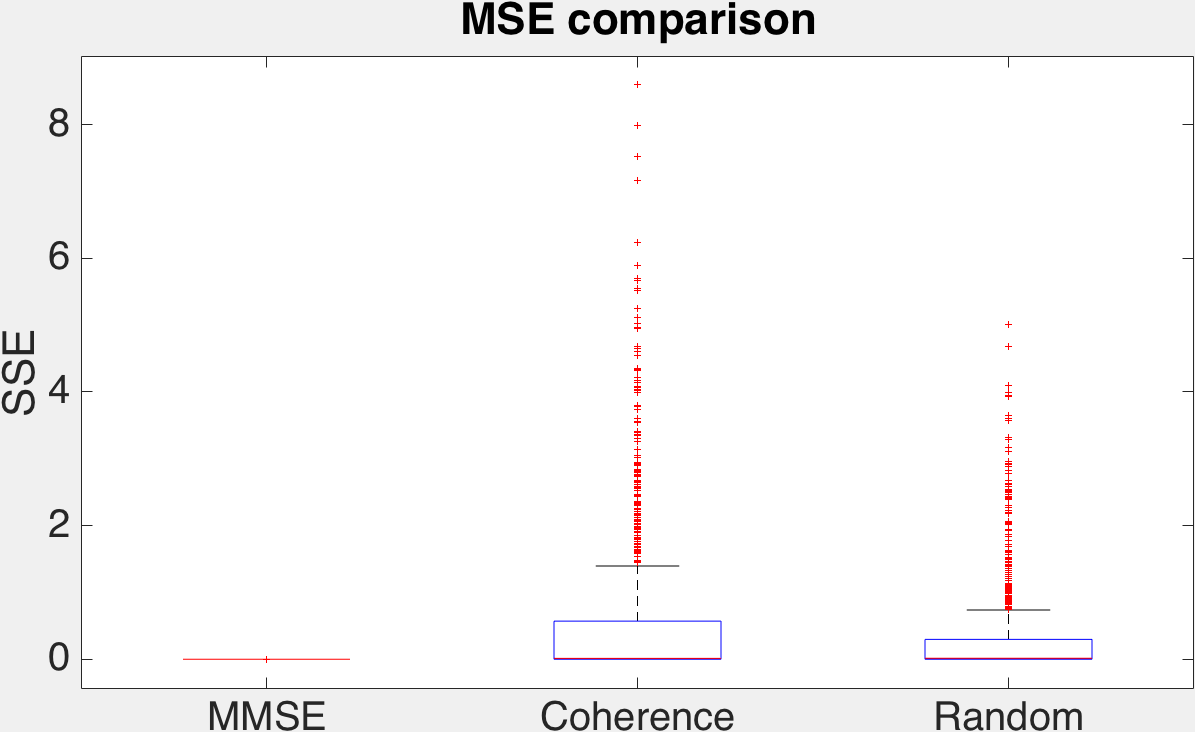}
    \caption{Comparison of sum of squared errors on a dataset of 250 sparse vectors, using $8 \times 8$ positive codes with $T=4$, optimized using the mean square error, optimized with coherence and a positive random code, targeted at sparsity of $0.12$}
    \label{fig:cactiMMSE4}
\end{figure}

\begin{figure}
    \centering
    \includegraphics[scale=0.2]{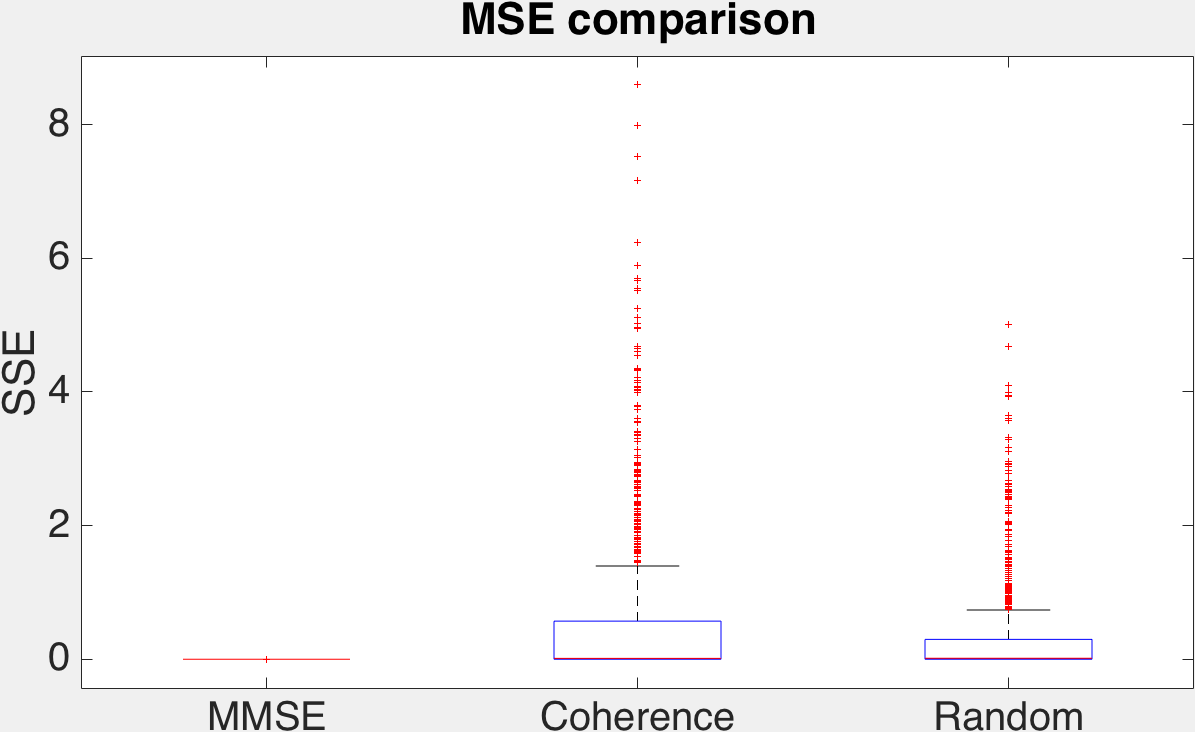}
    \caption{Comparison of sum of squared errors on a dataset of 250 sparse vectors, using $8 \times 8$ positive codes with $T=6$, optimized using the mean square error, optimized with coherence and a positive random code, targeted at sparsity of $0.08$}
    \label{fig:cactiMMSE6}
\end{figure}

Similarly, we design $8 \times 8$ codes for $T = $ 2, 4 and 6 (Figs.~\ref{fig:cactiMMSE2}, \ref{fig:cactiMMSE4} and \ref{fig:cactiMMSE6} respectively) targeted at perfectly sparse vectors with sparsity of $s = $ 0.2, 0.12 and 0.08 respectively. Optimizing the code using a random dataset of vectors drawn from similarly generated set of support sets, we compare the performance of the designed matrix on an independent test dataset of vectors having these supports.

The fact that mean squared error design does better than coherence design and random codes, where coherence design did worse than random codes, demonstrates clearly the potency of the MSE-based design over coherence-based design. It turns out, interestingly, that if we design matrices without the support set constraint with the same number of train vectors, MSE-based design still beats coherence and random by the same margin as above.

\section{Conclusion} \label{sec:conclusion}
The success of previous work using coherence to optimize sensing matrices served as a push for us to use coherence in the CACTI camera as well. This usage acquainted us with a pitfall in compressed sensing design using measures of matrix `quality'. The recovery error seems to be so loosely bounded, in this case, by both the RIC and the coherence, that optimizing the error bound in terms of the sensing matrix seems to do little towards optimizing the actual error. 

We then used the $\ell_1/\ell_\infty$ based criterion for perfectly sparse vectors and checked for the tightness of the bounds thus provided. We, however, discovered that the bound produced by this criterion, while indeed applicable to a broader set of signals than coherence, still produces fairly loose bounds.

We then resorted to calculating the RIC error bound in \cite{Cai2010} for reasonably-sized matrices at low sparsity levels, and compared the actual error over a dataset of sparse vectors to the bound. This lead us to the conclusion that even the RIC does not establish a tight bound on the recovery error.

A bit of thought revealed that all worst case bounds have a fundamental flaw: they take into consideration all vectors, and possibly a small, low-probability set of vectors loosen the bound up for all other vectors. We therefore proposed an average case analysis and presented a proof-of-concept design using this criterion.

\textbf{Reproducible Research:} All code used in generating results in this paper lives in the \texttt{src/descent-cacti}, \texttt{src/proof-comparison}, \texttt{src/ric-comparison} and \texttt{src/descent-mmse} directories in the Bitbucket repository at \href{https://bitbucket.org/alankarkotwal/coded-sourcesep}{\texttt{alankarkotwal/coded-sourcesep}}~\cite{Implement}. 

\bibliographystyle{IEEEtran}
\bibliography{references}

\end{document}